\documentclass[twocolumn,aps,superscriptaddress,floatfix]{revtex4-2}
\usepackage{amsmath}
\usepackage{amssymb}
\usepackage{graphicx}
\usepackage{color}
\usepackage{multirow}
\usepackage{siunitx}
\newcommand{\bea}{\begin{eqnarray}}
\newcommand{\eea}{\end{eqnarray}}
\newcommand{\bse}{\begin{subequations}}
\newcommand{\ese}{\end{subequations}}
\newcommand{\emb}{EuMg$_2$Bi$_2$}
\newcommand{\ems}{EuMg$_2$Sb$_2$}

\newcommand{\rFig}[1]{Figure~\ref{#1}}
\newcommand{\rtbl}[1]{Table~\ref{#1}}
\newcommand{\mub}{\mu_\text{B}}
\newcommand{\kso}{K_\text{so}}
\newcommand{\kd}{K_\text{D}}

\newcommand{\req}[1]{Eq.~(\ref{#1})}

\begin{document}

\title{Magnetic-field-induced $ab$-plane rotation of the Eu magnetic moments in trigonal EuMg$_2$Bi$_2$ and EuMg$_2$Sb$_2$ single crystals below their N\'eel temperatures}

\author{Santanu Pakhira}
\affiliation{Ames Laboratory, Iowa State University, Ames, Iowa 50011, USA}
\author{Yongbin Lee}
\affiliation{Ames Laboratory, Iowa State University, Ames, Iowa 50011, USA}
\author{Liqin Ke}
\affiliation{Ames Laboratory, Iowa State University, Ames, Iowa 50011, USA}
\author{D. C. Johnston}
\affiliation{Ames Laboratory, Iowa State University, Ames, Iowa 50011, USA}
\affiliation{Department of Physics and Astronomy, Iowa State University, Ames, Iowa 50011, USA}

\date{\today}

\begin{abstract}

The thermodynamic and electronic-transport properties of trigonal \emb\ in $ab$-plane magnetic fields ${\bf H}_x$ and the A-type antiferromagnetic structure have recently been reported. The Eu magnetic moments with spin $S = 7/2$ remain locked in the $ab$~plane up to and above the $ab$-plane critical field ${\bf H}_x^{\rm c} = 27.5$~kOe at which the Eu moments become parallel to ${\bf H}_x$.  Here additional measurements at low fields are reported that reveal a new spin-reorientation transition at a field $H_{c1} \approx 465$~Oe where the Eu moments remain in the $ab$~plane but become perpendicular to $H_x$.  At higher fields, the moments cant towards the field resulting in $M\propto H_x$ up to ${\bf H}_x^{\rm c}$.  Similar results are reported from measurements of the magnetic properties of \ems\ single crystals, where $H_{\rm c1}\approx 220$~Oe is found.  Theory is formulated that models the low-field magnetic behavior of both materials, and the associated anisotropies are calculated.  The $ab$-plane trigonal anisotropy in \ems\ is found to be significantly smaller than in \emb.

\end{abstract}

\maketitle

% The data for the plots were obtained from "AFM Domains with Anisotropy in a Field v8", 5/6/22

\section{Introduction}
\vspace{-0.1in}
The topological band properties have recently been investigated in crystallographically-ordered stoichiometric compounds because these properties are not as affected by disorder that occurs in substitutional solid solutions.  Of particular interest here are  the trigonal compounds \emb\ and \ems\ \cite{Wartenberg2002, Marshall2021} that can be cleanly grown as single crystals from self-fluxes~\cite{May2011}.   The bands in \emb\ have been studied~\cite{Kabir2019}, but not below the antiferromagnetic (AFM) ordering temperature $T_{\rm N}\approx 7$~K~\cite{May2011} of this compound.  The topological band structures of the nonmagnetic analogues YbMg$_2$Bi$_2$ and CaMg$_2$Bi$_2$ have also been investigated, where they are found to host topological surface states near the Fermi energy~\cite{Kundu2022}.

The low-temperature properties of \emb\ have been studied in detail~\cite{May2011, Pakhira2020, Pakhira2021}.  The magnetic structure below $T_{\rm N} = 6.7(1)$~K is an A-type AFM structure, in which the Eu moments with spin $S=7/2$ and spectroscopic splitting factor $g=2$ are ferromagnetically aligned parallel to an Eu layer of the structure, whereas the moments in adjacent layers along the $c$~axis are aligned antiparallel.  The specific heat $C_{\rm p}$ exhibits a $\lambda$-type anomaly at $T_{\rm N}$ and short-range magnetic order at temperatures~$T$ above $T_{\rm N}$.  The magnetic entropy $\approx R\ln(8)$ at $T\gg T_{\rm N}$ is consistent with that expected for Eu$^{2+}$ magnetic moments with magnitude $\mu=gS\,\mu_{\rm B} = 7\,\mu_{\rm B}$ with $g=2$ and spin $S= 7/2$, where $\mu_{\rm B}$ is the Bohr magneton.  The \mbox{$ab$-plane} electrical resistivity $\rho_{ab}(T)$ exhibits metallic character with a mild and disorder-sensitive upturn below $\sim 23$~K\@.

The crystal structure and magnetic properties of polycrystalline \ems\ have also been studied~\cite{Wartenberg2002}.  This compound contains Eu$^{2+}$ ions that exhibit AFM order below \mbox{$T_{\rm N}=8.2(3)$~K} with Curie-Weiss behavior in the magnetic susceptibility $\chi$ above $T_{\rm N}$.  The $^{151}$Eu M\"ossbauer spectra are consistent with this Eu$^{2+}$ oxidation state.  The band structure of \ems\ was also reported.  Detailed measurements of the magnetic and thermal properties of \ems\ single crystals have also been carried out recently~\cite{Pakhira2022}.  The $\rho(T)$ and angle-resolved photoemission (ARPES) measurements revealed semiconducting behavior with an energy gap of 370~meV\@.  The Eu$^{2+}$ magnetic moments exhibit A-type AFM order below \mbox{$T_{\rm N} = 8.0(2)$~K}\@.  An additional second-order transition of unknown origin occurs in the $\chi_{ab}(T)$ data at 3.0~K\@.

\emb\ and \ems\ crystals showed puzzling features in the field-dependent $ab$-plane magnetic susceptibility~$\chi_{ab}(T)$ below $T_{\rm N}$.   For \emb, The $\chi(T<T_{\rm N})$ measurements for $H\parallel ab$ and $H\parallel c$ in a field of 1~kOe were the same (isotropic)~\cite{May2011}, whereas for A-type AFM order one instead expects from molecular-field theory (MFT) that $\chi_{ab}(T\to0)=\chi(T_{\rm N})/2$ and $\chi_c(T\to0)=\chi(T_{\rm N})$~\cite{Johnston2012, Johnston2015, Johnston2021}.  The authors of Ref.~\cite{May2011} suggested that this observed behavior arises from $c$-axis magnetic ordering instead of the above A-type $ab$-plane ordering subsequently reported~\cite{Pakhira2021}.  Similarly, the data for $\chi_{ab}(T\leq T_{\rm N})$ in Ref.~\cite{Pakhira2020} for $H=0.5$--30~kOe were nearly independent of $T$\@.  These authors suggested that this $T$-independent $\chi_{ab}(T\leq T_{\rm N})$ behavior might arise from a field-induced helical magnetic structure with a turn angle of $\approx 120^\circ$ based on MFT~\cite{Johnston2012}, which we now know is not the correct magnetic structure~\cite{Pakhira2021}.  A similar anomalous behavior of $\chi_{ab}(H,T)$ was recently observed for \ems\ crystals~\cite{Pakhira2022}.

The above behavior of $\chi_{ab}(T\leq T_{\rm N})$ is correlated with anomalous low-field $M_{ab}(H)$ behavior for both \emb\ and \ems.  For these materials, these data show an unexpected positive curvature as shown in Fig.~\ref{Fig:EuMg2(Bi,Sb)2_MH}~\cite{Pakhira2021, Pakhira2022}.  It was suggested that this behavior may be caused by a field-induced reorientation of the magnetic moments in the three trigonal domains dictated by the crystal symmetry rather than by a field-induced change in the magnetic structure~\cite{Pakhira2021}.  Here we test this hypothesis and find strong evidence for it using a new  theory for magnetic-moment reorientation within the three trigonal domains in an $ab$-plane magnetic field.

\begin{figure}
\includegraphics [width=3.3in]{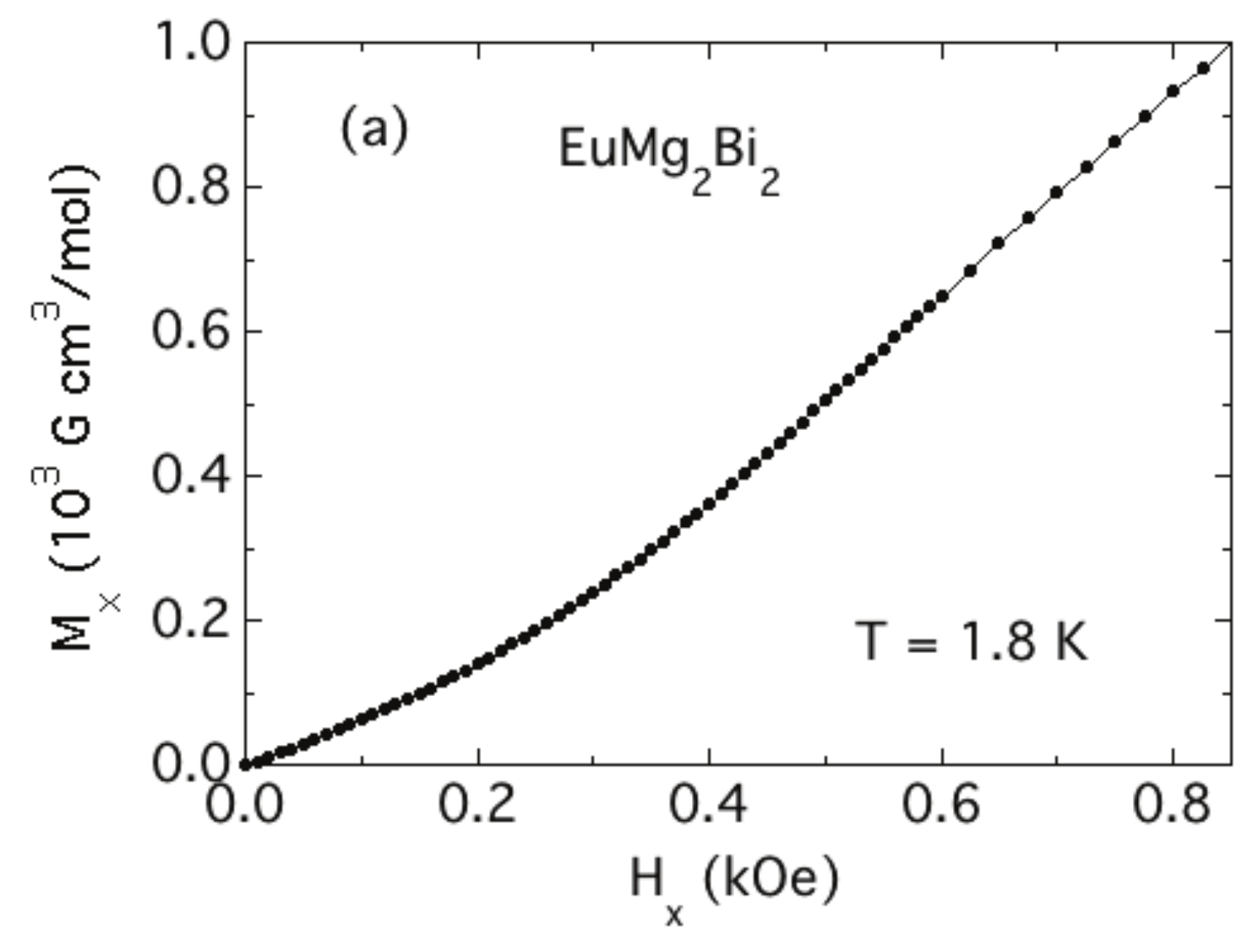}
\includegraphics [width=3.3in]{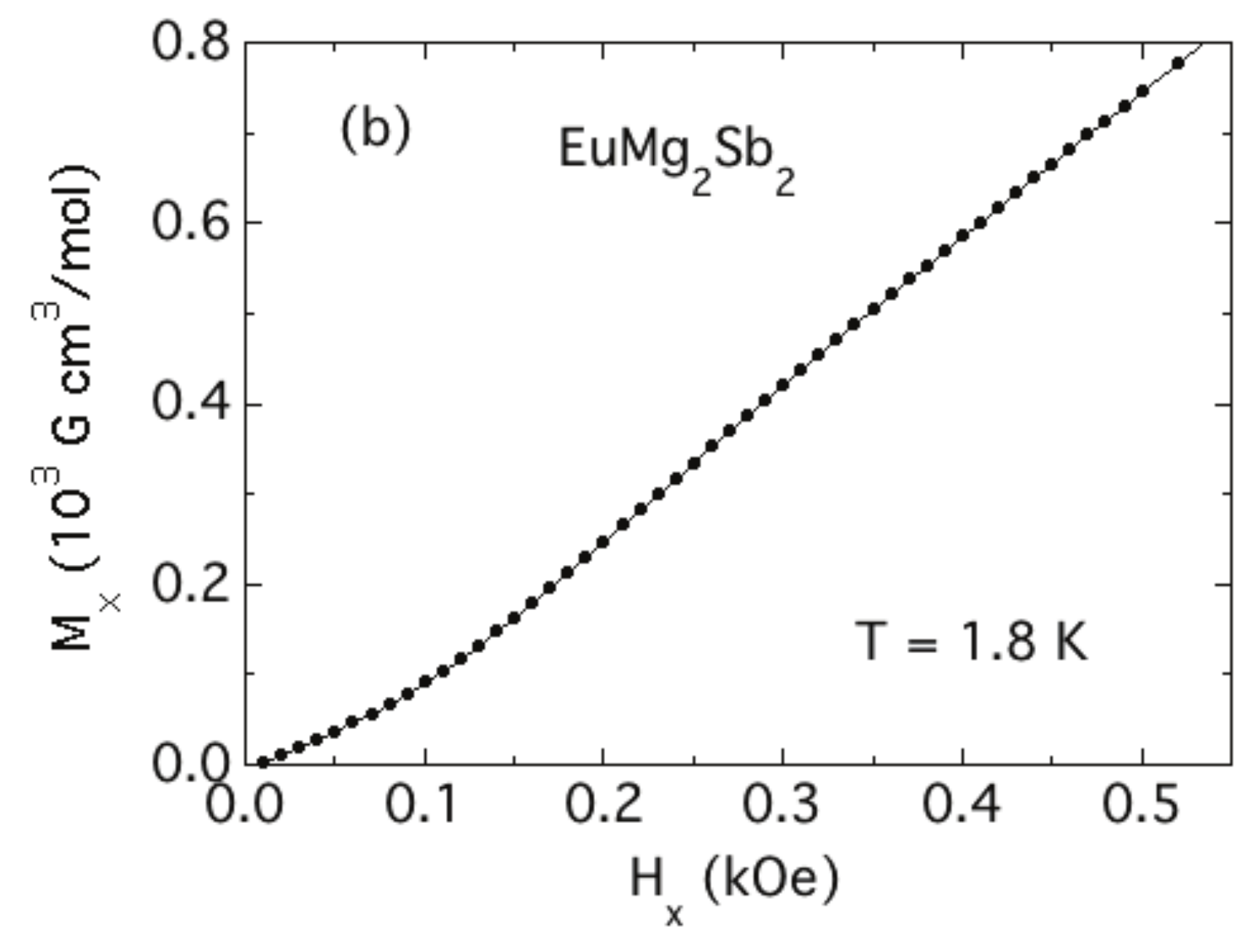}
\caption {Magnetization $M$ versus magnetic field $H_x$ applied in the $ab$~plane of single-crystal (a)~\emb\ and (b)~\ems. A distinct positive curvature is observed for \emb\ below $\sim 500$~Oe and for \ems\ below $\sim 200$~Oe~\cite{Pakhira2021, Pakhira2022}.}
\label{Fig:EuMg2(Bi,Sb)2_MH}
\end{figure}

The experimental and theoretical details are given in the following Sec.~\ref{Details}.
The theory for fitting the low-field $M_{ab}(H)$ data in Fig.~\ref{Fig:EuMg2(Bi,Sb)2_MH} is presented in Sec.~\ref{Theory}.
Fits to the data by the theory and calculations of the magnetic-dipole and magnetocrystalline anisotropies and comparisons with the fitted values are given in Sec.~\ref{Fits}.
Concluding remarks are given in Sec.~\ref{Conclusion}.
\vspace{-0.2in}

\section{\label{Details} Experimental and Theoretical Details}
\vspace{-0.1in}
Trigonal \emb\ single crystals with room-temperature hexagonal lattice parameters \mbox{$a = 4.7724(3)$~\AA\ }  and $c = 7.8483(5)$~\AA\ were grown by a self-flux method with starting composition EuMg$_4$Bi$_6$ as reported earlier~\cite{Pakhira2020, Pakhira2021}. \ems\ single crystals with room-temperature hexagonal lattice parameters $a = b = 4.6861(3)$~\AA\ and $c = 7.7231(5)$~\AA\ were also grown using self-flux with starting composition EuMg$_4$Sb$_{16}$~\cite{Pakhira2022}. The magnetization measurements were carried out using a Magnetic Properties Measurement System (MPMS, Quantum Design, Inc.)\@ operating in the temperature range 1.8--300~K and with magnetic fields~$H$ up to 5.5 T \mbox{(1~T = 10$^4$~Oe)}.  For measurements with the field applied  in the $ab$~plane,  the magnetic field was perpendicular to the [100] direction of the hexagonal unit cell as determined from Laue x-ray diffraction measurements, whereas in Ref.~\cite{Pakhira2020} the field was applied in an arbitrary direction in the $ab$~plane.

Two contributions to the magnetic anisotropy (MA), magnetic-dipole anisotropy (MDA) and magnetocrystalline anisotropy (MCA), are calculated and compared to experiments. 
MDA energies (MDAE) $\kd$ are calculated by a direct lattice summation over Eu atoms in a sufficiently large sphere to ensure convergence.
MCA energies (MCAE) $K$ are calculated using density functional theory (DFT) as implemented in Vienna \emph{ab initio} simulation package (\textsc{Vasp})~\cite{kresse1993prb, kresse1996prb}.
We calculate the total energies of AFM A-type and FM orderings with different spin orientations.
The MCAE of a particular magnetic ordering is calculated as
\bea
K=E_{001}-E_{100},
\label{eq:kc}
\eea
where $E_{001}$ and $E_{100}$ are the total energies for the magnetization oriented along the $[001]$ and $[100]$ directions, respectively.
Positive (negative) $K$ corresponds to easy-plane (easy-axis) anisotropy.
To decompose the MCAE, we also evaluate the on-site SOC energy $\langle V_\text{so} \rangle$ and the corresponding anisotropy
\bea
\kso=\frac{1}{2}\langle V_\text{so} \rangle_{001}{-}\frac{1}{2}\langle V_\text{so} \rangle_{100}.
\label{eq:kso}
\eea
As shown later, we found $\kso\approx K$, which is expected from the second-order perturbation theory~\cite{antropov2014ssc, ke2015prb}.
However, unlike $K$, $\kso$ can be decomposed into sites, spins, and orbital pairs, providing an understanding of the MCA mechanism in a system~\cite{antropov2014ssc, ke2015prb, ke2019prb, ke2016prbA}.

The experimental crystal structures~\cite{Pakhira2020,Pakhira2022} are used in all calculations.
The nuclei and core electrons are described by the projector augmented-wave potential~\cite{kresse1999prb}, and the wave functions of valence electrons are expanded in a plane-wave basis set with a cutoff energy of up to \SI{520}{eV}.
The spin-orbit coupling (SOC) is included using the second-variation procedure~\cite{koelling1977jpcs,shick1997prb}.
In addition, the Hubbard Coulomb interaction $U=\SI{6}{\eV}$ is included for better accounting for the strong correlation of Eu-$4f$ electrons~\cite{dudarev1998prb}.
The $k$-point integration was performed using a modified tetrahedron method with Bl\"ochl corrections.
A $16\times 16\times 4$ $k$-point mesh is used for MCAE calculations to ensure sufficient convergence.

\vspace{-0.2in}
\section{\label{Theory} Theory}
\vspace{-0.1in}
\subsection{Overview}

\begin{figure}
\includegraphics [width=3.3in]{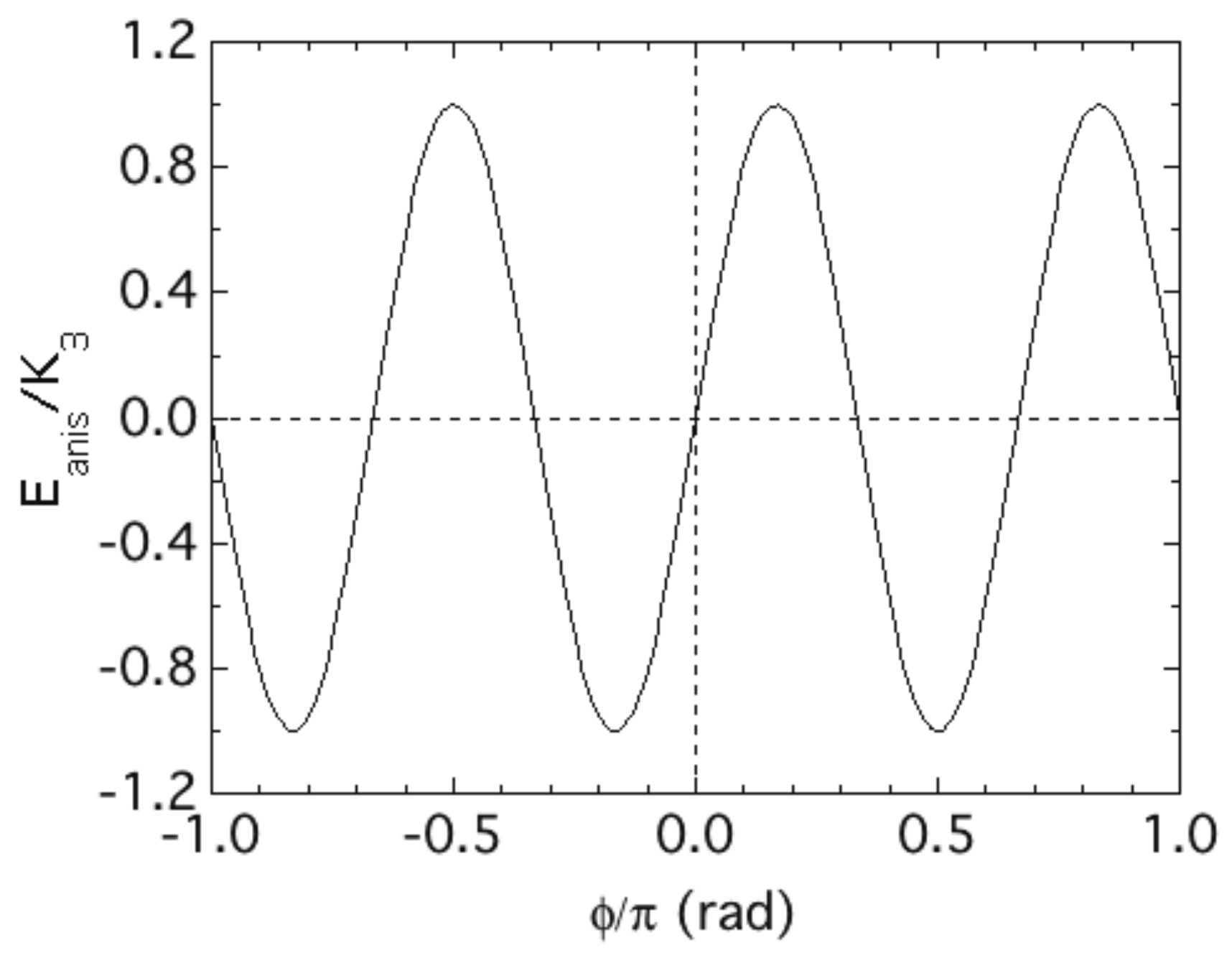}
\caption {In-plane trigonal anisotropy energy $E_{\rm anis}$ normalized by the anisotropy constant $K_3$ of a moment versus its angle $\phi$ with respect to the positive $x$~axis.  The three minima of $E_{\rm anis}$ are at \mbox{$\phi/\pi = -5/6,\ -1/6$, and 1/2.}}
\label{Fig:AnisEnergyVsPhi}
\end{figure}

\begin{figure}[ht]
\includegraphics [width=2.3in]{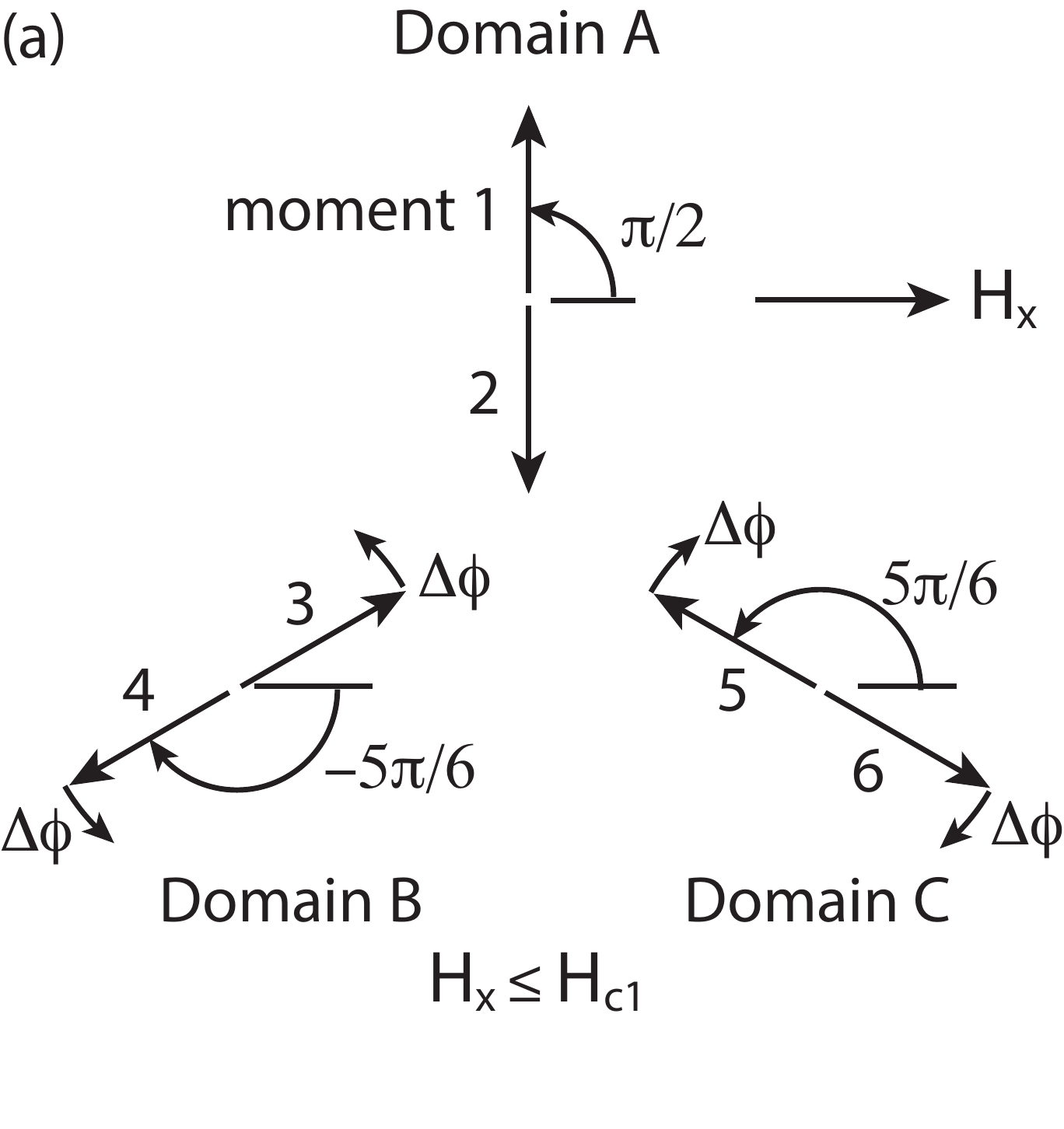}
\includegraphics [width=2.in]{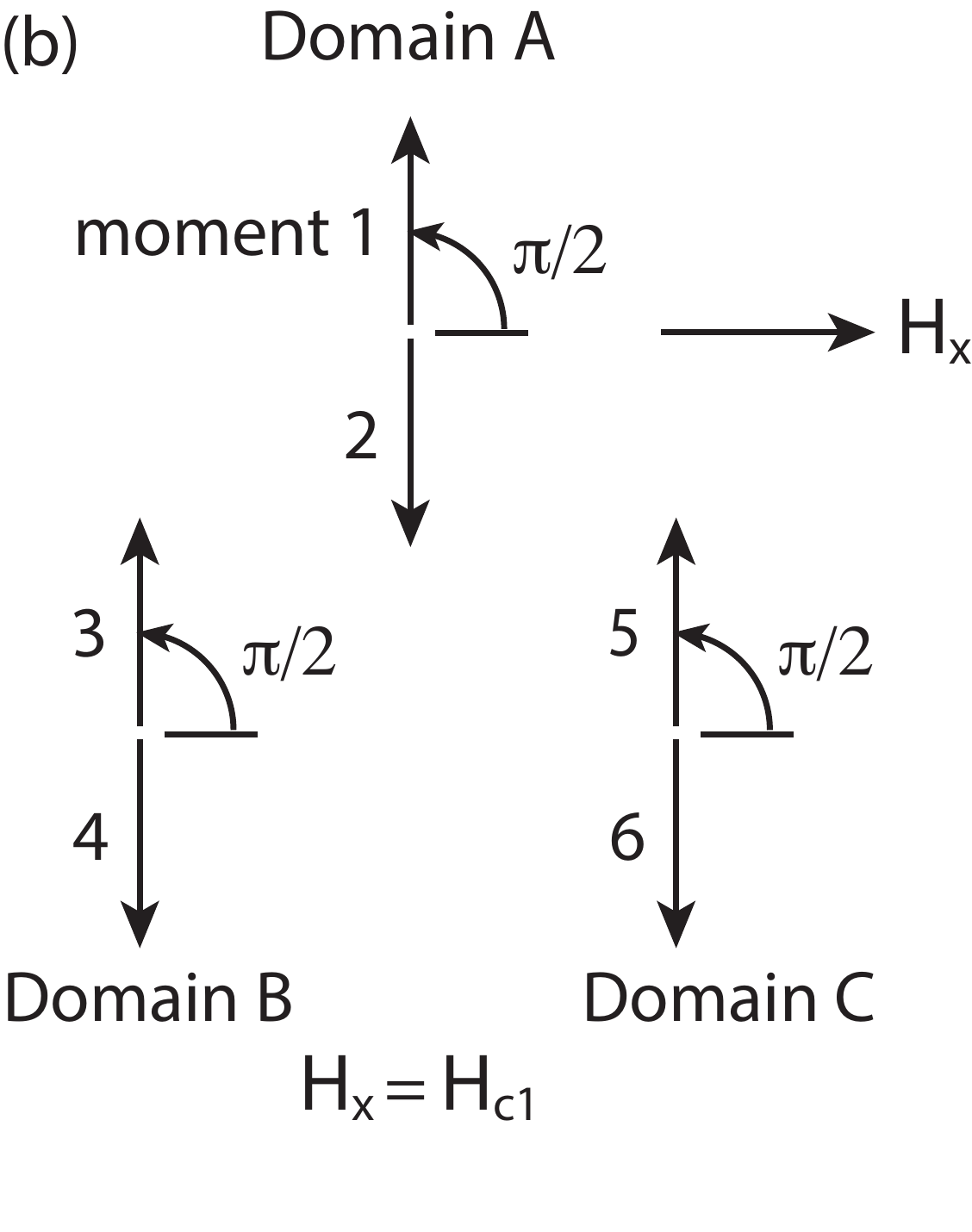}
\includegraphics [width=2.in]{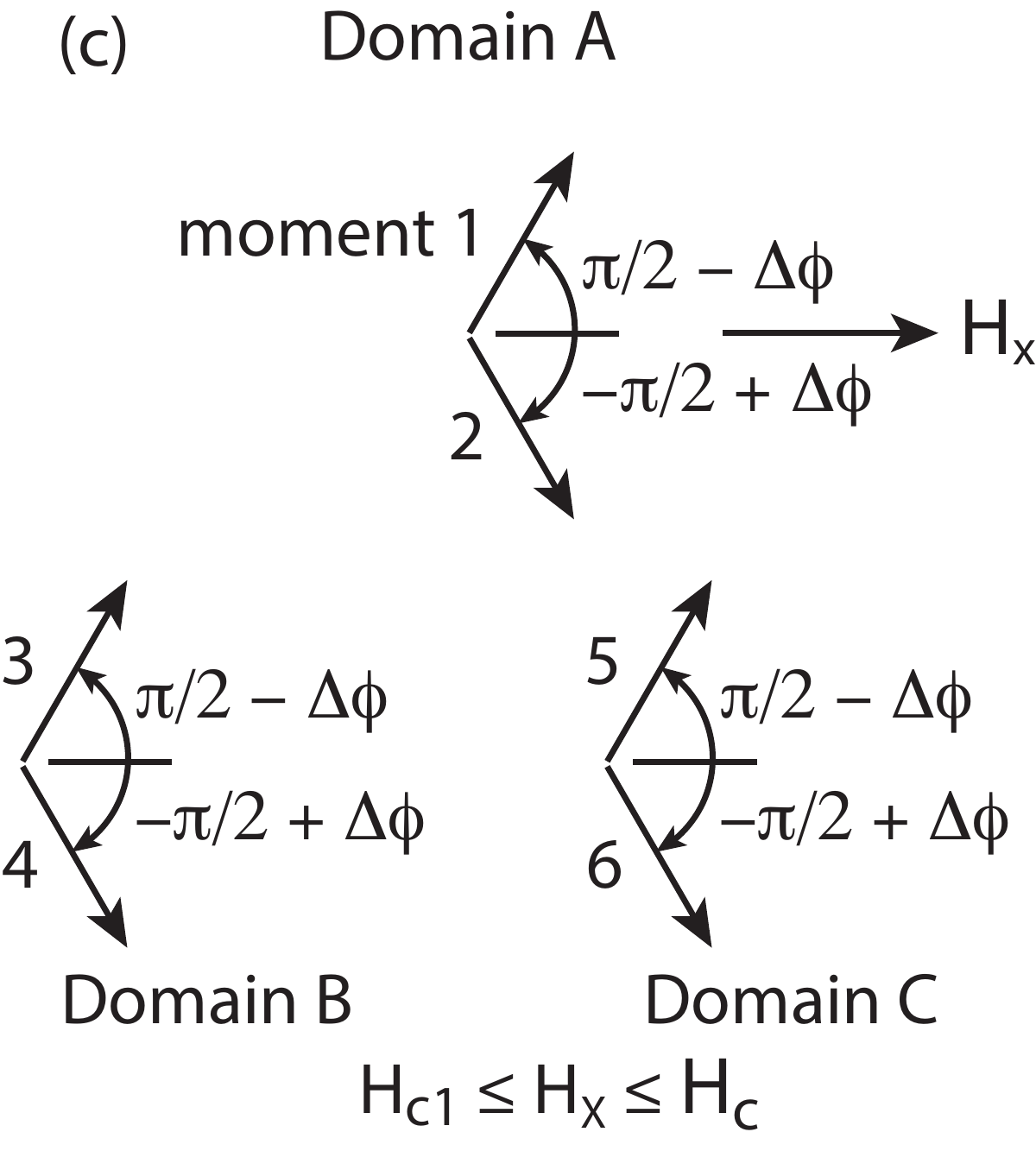}
\caption {(a)~Reorientation of the Eu magnetic moments in the three trigonal antiferromagnetic domains in a small $ab$-plane magnetic field $H_x<H_{\rm c1}$. The arrows indicate the direction and increment $\Delta\phi$ of rotation of the moments in domains B and~C towards the vertical orientation.  The moments in each domain remain antiparallel to each other for $H_x<H_{\rm c1}$ apart from a   small canting towards the magnetic field direction ($\lesssim 1^\circ$) that gives rise to the measured magnetization. (b)~Orientation of the moments at the critical field $H_x=H_{\rm c1}$ where all moments are perpendicular to $H_x$ except for the small canting towards $H_x$.  (c)~Canting of all moments toward $H_x$ for \mbox{$H_{\rm c1} < H_x < H_{\rm c}$.}  At the critical field $H_c$ all moments are aligned ferromagnetically in the direction of  ${\bf H}_x$.}
\label{Fig:TrigDomains}
\end{figure}

As discussed later, \emb\ and \ems\ have a strong XY anisotropy which keeps the ordered moments confined to the trigonal $ab$ plane for magnetic field applied parallel to this plane.  The $ab$-plane trigonal anisotropy energy has the form
\bea
E_{\rm anis} = K_3 \sin(3\phi),
\label{Eq:Eanis}
\eea
where $K_3$ is the positive trigonal anisotropy constant and $\phi$ is the angle of a magnetic moment with respect to the positive $x$~axis, which is the direction of the applied field ${\bf H}_x$.  A plot of $E_{\rm anis}$ normalized by $K_3$ versus  $\phi/\pi$ is shown in Fig.~\ref{Fig:AnisEnergyVsPhi}, where the angles~$\phi$  of the negative anisotropy-energy minima with respect to the in-plane $x$~axis are given in the figure caption.  In $H=0$, these are the angles of the moments in each domain as illustrated in Fig.~\ref{Fig:TrigDomains}(a) with $\Delta\phi = 0$.  In $H=0$, each domain contains two moments at 180$^\circ$ with respect to each other, where the two moments are respectively in alternating layers along the $c$~axis of the A-type AFM structure.  As shown, moments 1 and~2 are in Domain~A, 3 and~4 are in Domain B, and 5 and~6 are in Domain~C\@.  The moments in different domains are assumed not to interact.

\begin{figure}
\includegraphics [width=3.in]{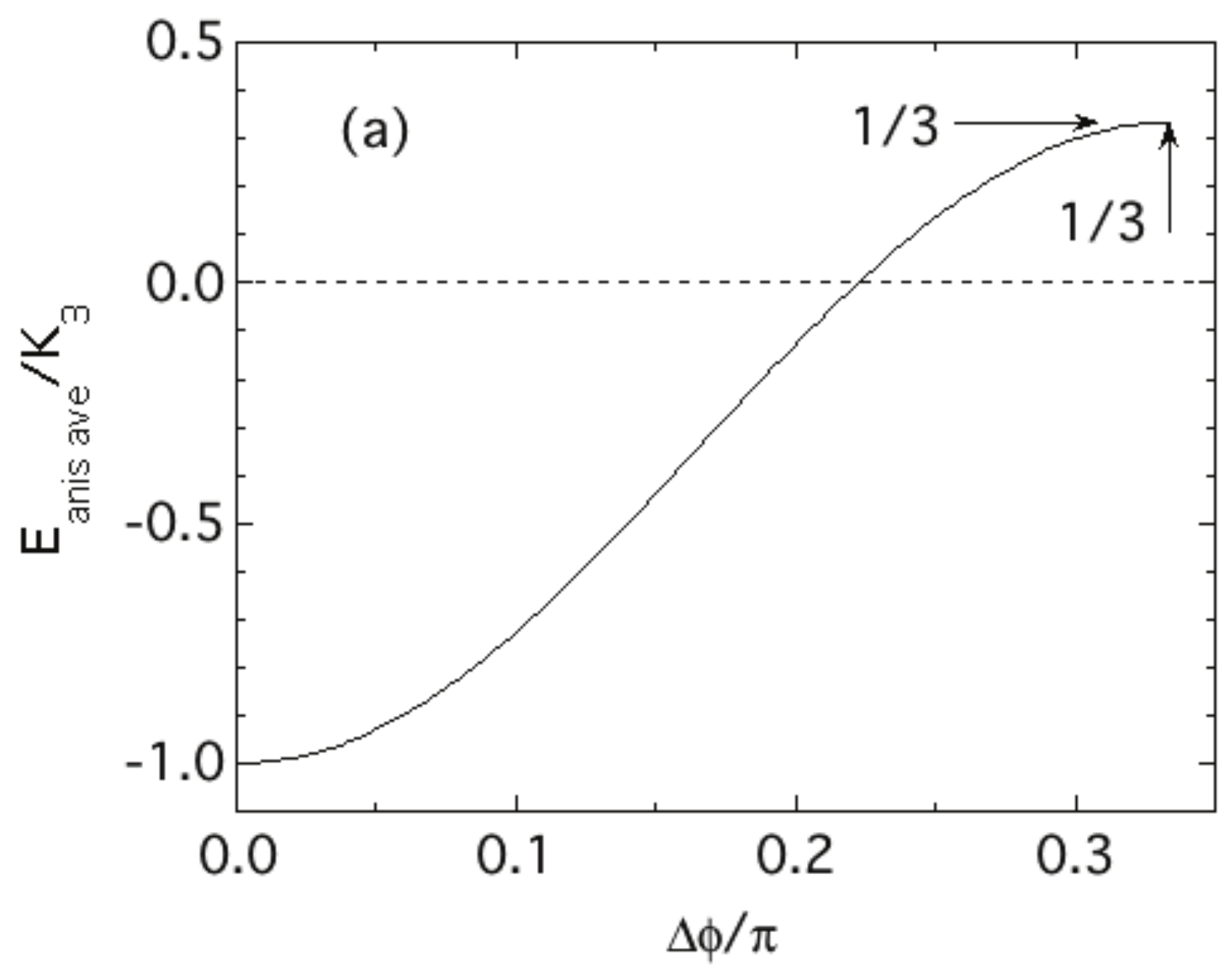}
\includegraphics [width=3.in]{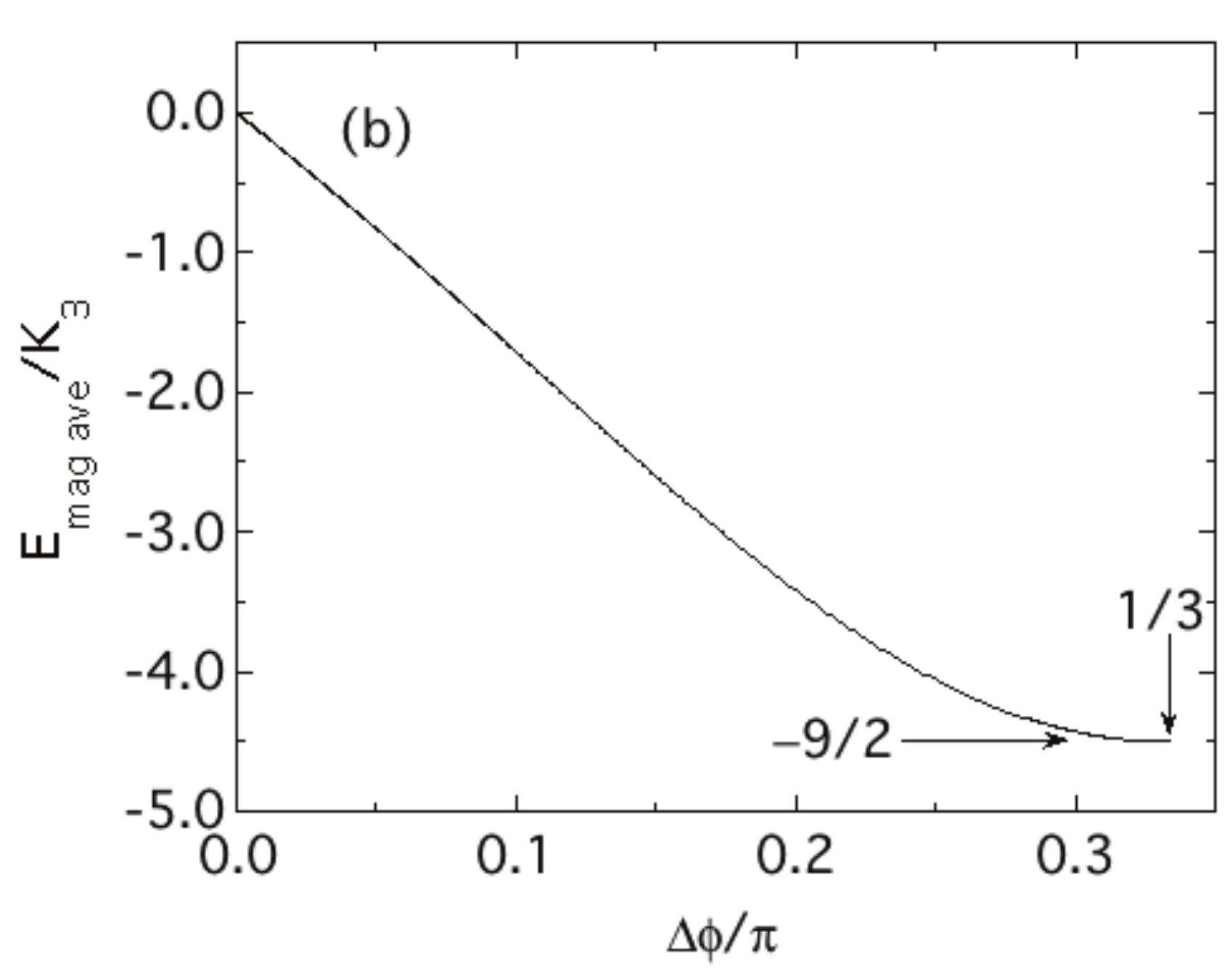}
\includegraphics [width=3.in]{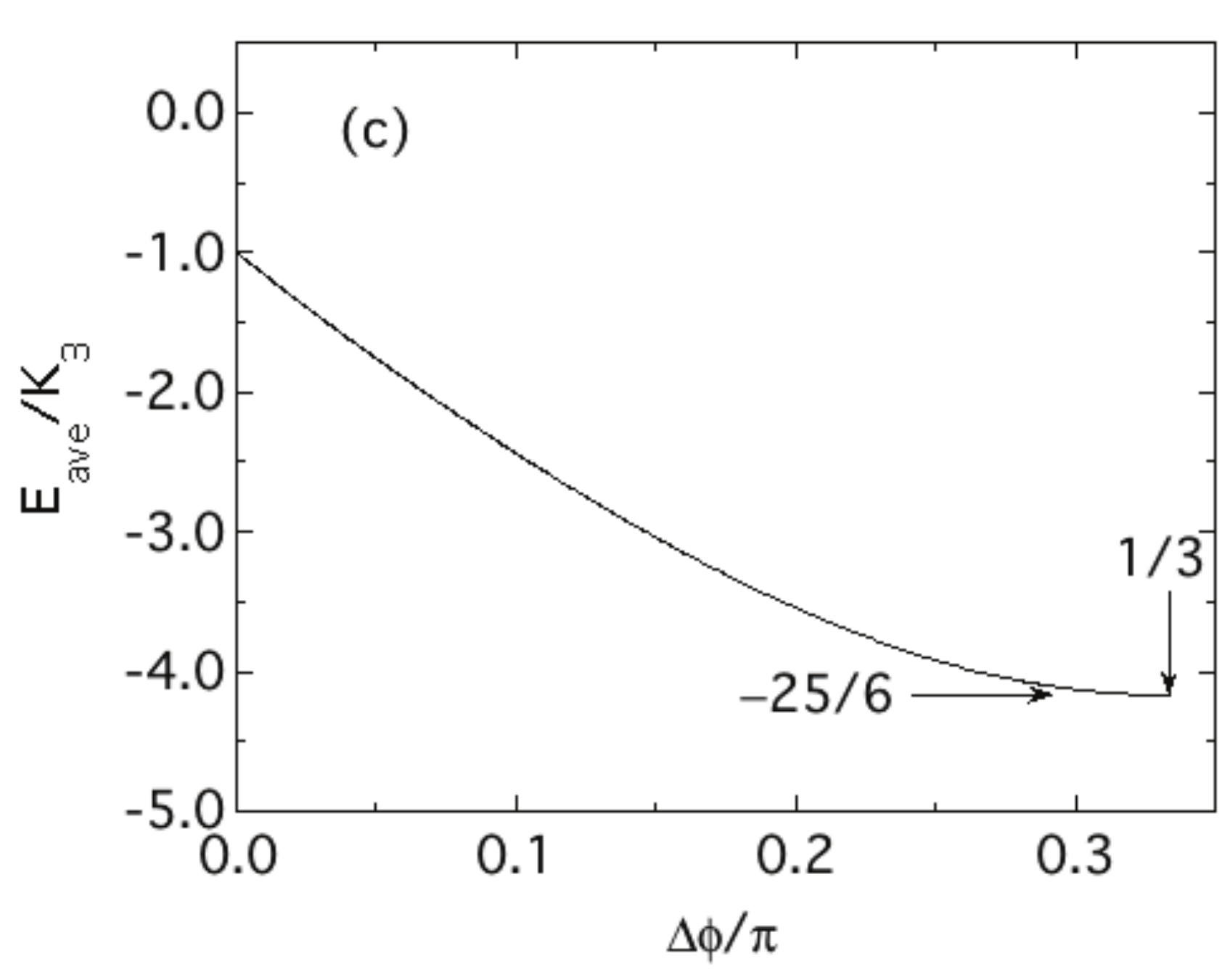}
\caption {(a)~Average anisotropy energy $E_{\rm anis\,ave}$, (b)~average magnetic energy $E_{\rm mag\,ave}$, and (c) the average total energy $E_{\rm ave}$, all normalized by the anisotropy constant $K_3$, versus the tilt angle $\Delta\phi/\pi$ of the AFM moments in Fig.~\ref{Fig:TrigDomains}(a) except for the moments in Domain~A for which $\Delta\phi = 0$.  The value $|\Delta\phi|/\pi = 1/3$ occurs at $H_x=H_{\rm c1}$.}
\label{Fig:Eanis_mag_Ave}
\end{figure}

%% \clearpage

There is a small field $H_x \equiv H_{\rm c1}$ at which the moments in all three domains become almost perpendicular to $H_x$ as shown in Fig.~\ref{Fig:TrigDomains}(b), where we eventually determine $H_{\rm c1} =  465$~Oe and 220~Oe for \emb\ and \ems, respectively (see~Fig.~\ref{Fig:EuMg2(Bi,Sb)2_MH}).  The tilt angle $\phi_{\rm tilt}$ in a field $H_x$ of each of the two moments in each domain at $H_{c1}$ is given by $\phi_{\rm tilt} = \arcsin(H_x/H^{\rm c}_{ab})$, where $H^{\rm c}_{ab}$ is the critical field in the $ab$ plane at which all moments become parallel to $H_x$.  Using  $H_{c1}= 465$~Oe and  $H^{\rm c}_{ab}= 27,500$~Oe for \emb, the tilt angle at $H_{\rm c1}$ is $\phi_{\rm tilt} = 0.75^\circ$ at \mbox{$H_x=H_{\rm c1}$} which gives rise to the $x$-axis magnetization observed at that field.  This tilt angle is negligible compared to the value of 90$^\circ$ at saturation.  Thus for fields less than $H_{\rm c1}$, we can consider the two moments in each domain to be (almost) locked at angles of $\approx 180^\circ$ to each other as shown by the arrows in Fig.~\ref{Fig:TrigDomains}(a).

The reason for the small value of $\phi_{\rm tilt}$ is that the anisotropy energy is much smaller than the exchange interaction energy between the two spins in each domain which tends to keep them aligned antiparallel.  For example, the $ab$-plane trigonal anisotropy parameter in \emb\ according to Eq.~(\ref{Eq:K3EMB}) below is \mbox{$K_3 = 6.5 \times 10^{-8}$~eV/Eu,} whereas the Heisenberg exchange interaction between a spin in one ferromagnetically-aligned layer and the spins aligned in the opposite direction in one of the two nearest layers is $J_1=1.97\times10^{-5}$\,eV~\cite{Pakhira2020}.

\subsection{$0\leq H_x\leq H_{\rm c1}$}

In the small fields $0 \leq H_x\leq H_{\rm c1}$, the angles of the locked moments in domains A, B, and C in Fig.~\ref{Fig:TrigDomains}(a) with respect to the positive $x$~axis are respectively given by
\bea
\phi_{\rm A} &=& \frac{\pi}{2},\nonumber\\
\phi_{\rm B} &=& -\frac{5\pi}{6} + \Delta\phi \quad (0\leq \Delta\phi \leq \pi/3), \label{Eqs:phiABC}\\
\phi_{\rm C} &=& -\frac{\pi}{6} - \Delta\phi.  \quad (0\leq \Delta\phi \leq \pi/3), \nonumber
\eea
where the magnitude of the angle of the moments in Domains~B and~C with respect to their initial angles is $\Delta \phi$.  This is not the angle $\phi_{\rm tilt}$ between the two locked moments in a domain discussed above.  The anisotropy energy averaged over the moments in the three domains in the field range $0\leq H_x\leq H_{\rm c1}$ using Eqs.~(\ref{Eq:Eanis}) and (\ref{Eqs:phiABC}) is
\bea
E_{\rm anis\,ave} &=& \frac{K_3}{3}[\sin(3\phi_{\rm A}) +  \sin(3\phi_{\rm B}) + \sin(3\phi_{\rm C})]\nonumber\\
&=& - \frac{K_3}{3}[1+2\cos(3\Delta\phi)].\label{Eq:EanisAve}
\eea
A plot of $E_{\rm anis\,ave}/K_3$ versus  $\Delta\phi$ is shown in Fig.~\ref{Fig:Eanis_mag_Ave}(a).  The negative (lowest-energy) value of $E_{\rm anis\,ave}/K_3$ at $\Delta\phi=0$ is seen to become positive with increasing $\Delta\phi$.

The magnetic energy in the regime $0\leq H_x \leq H_{\rm c1}$ is given by
\bse
\bea
E_{\rm mag} &=& -M_xH_x = -[\chi_\perp H_x \sin(\phi)] H_x \label{Eq:Emag}\\
&=& -\chi_\perp H_x^2\sin(\phi),\nonumber
\eea
where $\chi_x\equiv \chi_\perp = M_x/H_x$ is the $ab$-plane magnetic susceptibility at $T=0$ when all moments are perpendicular to ${\bf H}_x$, {\it i.e.}, when $\phi=\pi/2$.  Summing over the angles of the moments in the three domains in Eq.~(\ref{Eqs:phiABC}) and dividing by $3$ gives the average magnetic energy as
\bea
E_{\rm mag\ ave} = -\frac{\chi_\perp H_x^2}{3}\left[1+2\sin^2\left(\frac{\pi}{6} +\Delta\phi\right) \right].
\label{Eq:EmagAve}
\eea
\ese
The total average energy is $E_{\rm ave} = E_{\rm anis\,ave} + E_{\rm mag\,ave}$, which is given by Eqs.~(\ref{Eq:EanisAve}) and (\ref{Eq:EmagAve}) as
\bse
\label{Eqs:Eave}
\bea
E_{\rm ave} &=& -\frac{K_3}{3}[1+2\cos(3\Delta\phi)] \label{Eq:Eave}\\  
&&-\frac{\chi_\perp H_x^2}{3}\left[1+2\sin^2\left(\frac{\pi}{6} +\Delta\phi\right) \right]\nonumber.
\eea

\clearpage   

\begin{figure}
\includegraphics [width=3.3in]{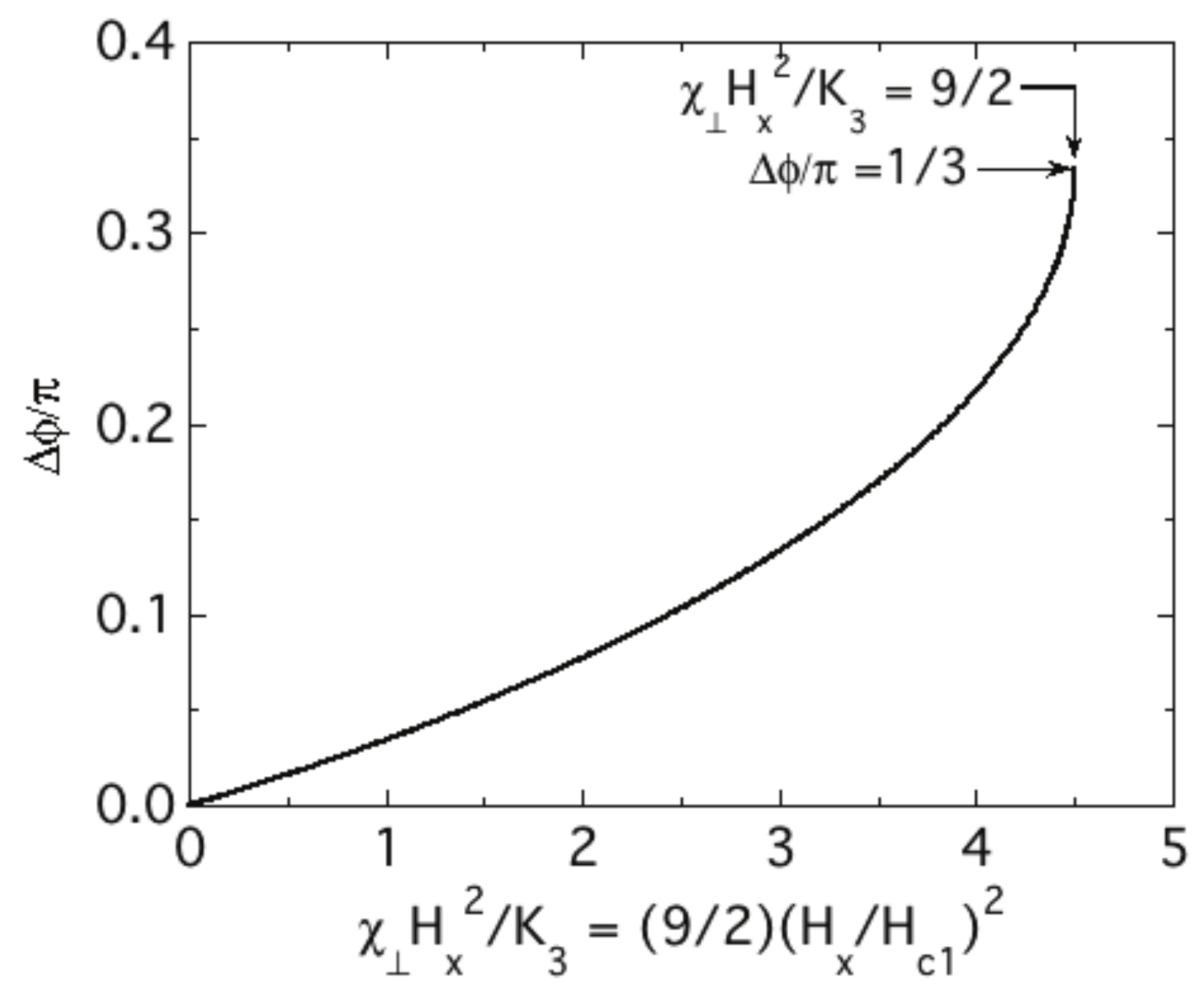}
\caption {The angle $\Delta\phi/\pi$ in Fig.~\ref{Fig:TrigDomains}(a) versus $\chi_\perp H_x^2/K_3$ in Eqs.~(\ref{Eqs:Eave}).   The alternate abscissa label is obtained using Eq.~(\ref{Eq:chiperponK3}).}
\label{Fig:DeltaPhiVSchiPerpHx2}
\end{figure}

In order to minimize $E_{\rm ave}$ with respect to $\Delta\phi$ at each value of $H_x$ with $0\leq H_x\leq H_{\rm c1}$, we first normalize the average energy in Eq.~(\ref{Eq:Eave}) by $K_3$, yielding
\bea
\frac{E_{\rm ave}}{K_3} &=& -\frac{1}{3}\bigg\{1+2\cos(3\Delta\phi)] \label{Eq:Eave2}\\  
&&+\frac{\chi_\perp}{K_3} H_x^2 \left[1+2\sin^2\left(\frac{\pi}{6} +\Delta\phi\right) \right]\bigg\},\nonumber
\eea
\ese
where both sides of the equation in cgs units are dimensionless since the cgs units of $\chi_\perp$ are cm$^3$, those of $H_x^2$ are erg/cm$^3$, and those of $K_3$ are ergs.  
Minimizing $E_{\rm ave}/K_3$ with respect to the quantity $\chi_\perp H_x^2/K_3$ yields $\Delta\phi$ versus $\chi_\perp H_x^2/K_3$ as plotted in Fig.~\ref{Fig:DeltaPhiVSchiPerpHx2}.  Then using the data in the figure for the last point on the right end of the plot at $H = H_{\rm c1}$ for which $\Delta\phi=\pi/3$, we obtain
\bse
\bea
\frac{\chi_\perp H_{\rm c1}^2}{K_3} = \frac{9}{2}.
\label{Eq:ChiPerpHc12}
\eea
We calculated the value of the constant on the right-hand side of Eq.~(\ref{Eq:ChiPerpHc12}) to high precision, and it appears to be exact.  Equation~(\ref{Eq:ChiPerpHc12}) gives
\bea
\frac{\chi_\perp}{K_3} = \frac{9/2}{H_{\rm c1}^2}.
\label{Eq:chiperponK3}
\eea
\ese
Inserting this expression into Eqs.~(\ref{Eq:EmagAve}) and~(\ref{Eq:Eave2}) respectively gives
\bse
\bea
\frac{E_{\rm mag\ ave}}{K_3} = -\frac{3}{2} \left(\frac{H_x}{H_{\rm c1}}\right)^2 \left[1+2\sin^2\left(\frac{\pi}{6} +\Delta\phi\right) \right]
\label{Eq:EmagAve2}
\eea
and
\bea
\frac{E_{\rm ave}}{K_3} &=& -\frac{1}{3}\bigg\{1+2\cos(3\Delta\phi) \label{Eq:Eave33}\\  
&&+ \frac{9}{2}\left(\frac{H_x}{H_{\rm c1}}\right)^2\left[1+2\sin^2\left(\frac{\pi}{6} +\Delta\phi\right) \right]\bigg\}.\nonumber
\eea
\ese

A plot of $E_{\rm mag\ ave}/K_3$ versus $\Delta\phi$ for $0\leq H_x \leq H_{\rm c1}$ is shown in Fig.~\ref{Fig:Eanis_mag_Ave}(b) and a corresponding plot of $E_{\rm ave}/K_3$ versus $\Delta\phi$  is shown in Fig.~\ref{Fig:Eanis_mag_Ave}(c).  For $E_{\rm ave}/K_3$, the minimum energy is obtained at $\Delta\phi = \pi/3$ for which the moments in all three domains are perpendicular to $H_x$ apart from the slight canting towards $H_x$ discussed above that gives rise to the observed magnetization versus $H_x$.

\begin{figure}
\includegraphics [width=3.3in]{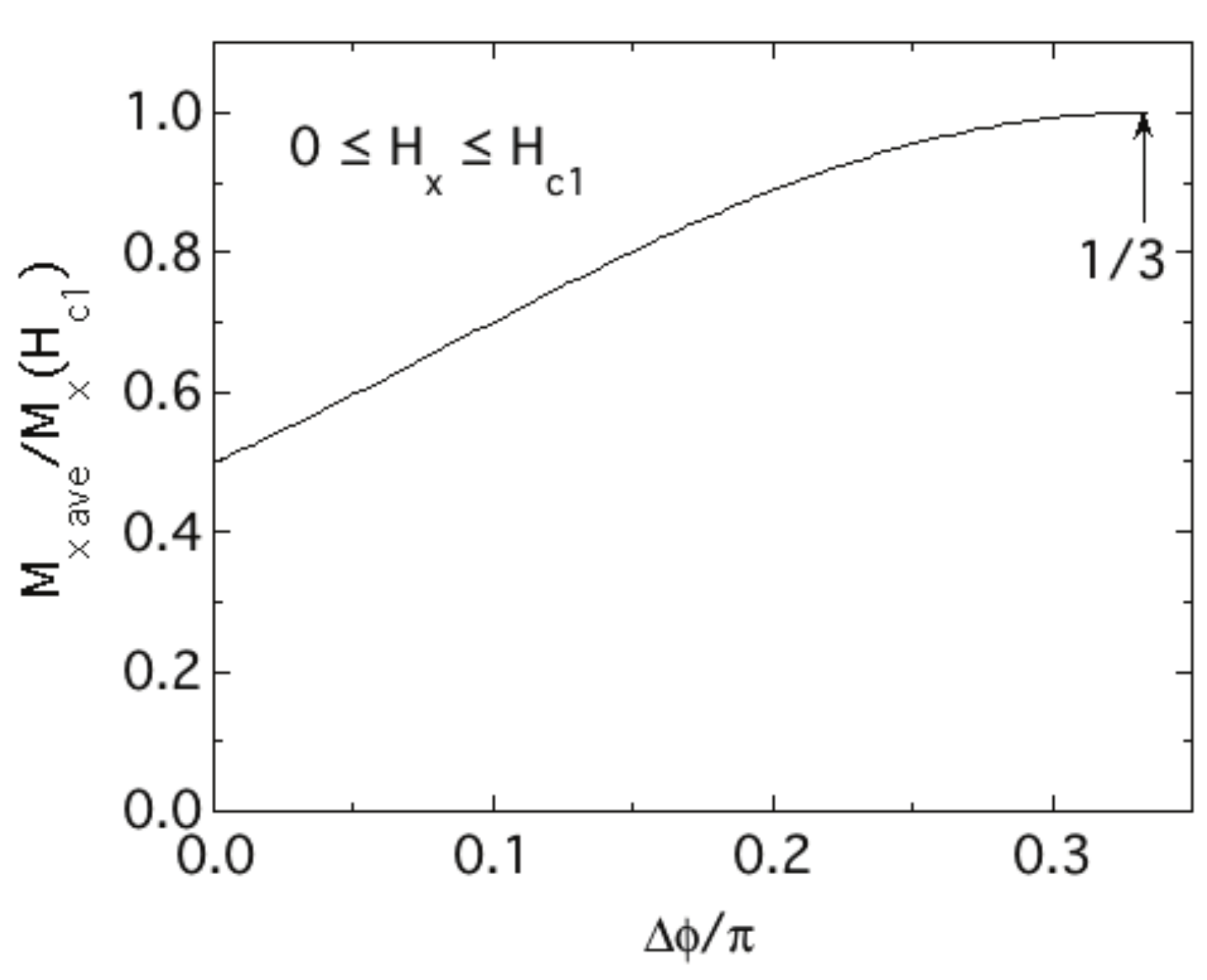}
\caption {Average magnetization $M_{x\,\rm ave}$ in the direction of the field $H_x$ normalized by $M_x(H_{\rm c1})$  versus the angle $\Delta\phi/\pi$ of the AFM domains in Fig.~\ref{Fig:TrigDomains}(a), where the angle $\Delta\phi/\pi=1/3$  corresponds to the field $H_x=H_{\rm c1}$.}
\label{Fig:MuxOnMuxHc1}
\end{figure}

For $0\leq H_x\leq H_{\rm c1}$,  the magnetization $M_x$ of the two collinear moments in a domain at $T=0$ versus $H_x$ only arises from the perpendicular component of {\bf M}, because the parallel component gives no contribution at $T=0$~K\@.  The normalized magnetization averaged over the three domains using Eqs.~(\ref{Eqs:phiABC}) is
\bea
\frac{M_{x\,\rm ave}}{M_x(H_{\rm c1})} &=& \frac{1}{3}\left[\sin^2(\phi_{\rm A})+\sin^2(\phi_{\rm B})+\sin^2(\phi_{\rm C})\right]\nonumber\\
&=& \frac{1}{3}\left[1+2\sin^2\left(\frac{\pi}{6}+\Delta\phi\right)\right].
\eea
Thus if $\Delta\phi=0$, $M_{x\,\rm ave}/M_x(H_{\rm c1})=1/2$, whereas if \mbox{$\Delta\phi=\pi/3$}, all the moments are perpendicular to $H_x$, giving the maximum value $M_{x\,\rm ave}/M_x(H_{\rm c1})=1$ for the locked moments in the three domains.  A plot of $M_{x\,\rm ave}/M_x(H_{\rm c1})$ versus $\Delta\phi/\pi$ over the relevant range $0\leq \Delta\phi/\pi\leq 1/3$ is shown in Fig.~\ref{Fig:MuxOnMuxHc1}.

%\clearpage

\subsection{$H_{\rm c1}\leq H_x\leq H^{\rm c}_{ab}$}

\begin{figure}
\includegraphics [width=3.in]{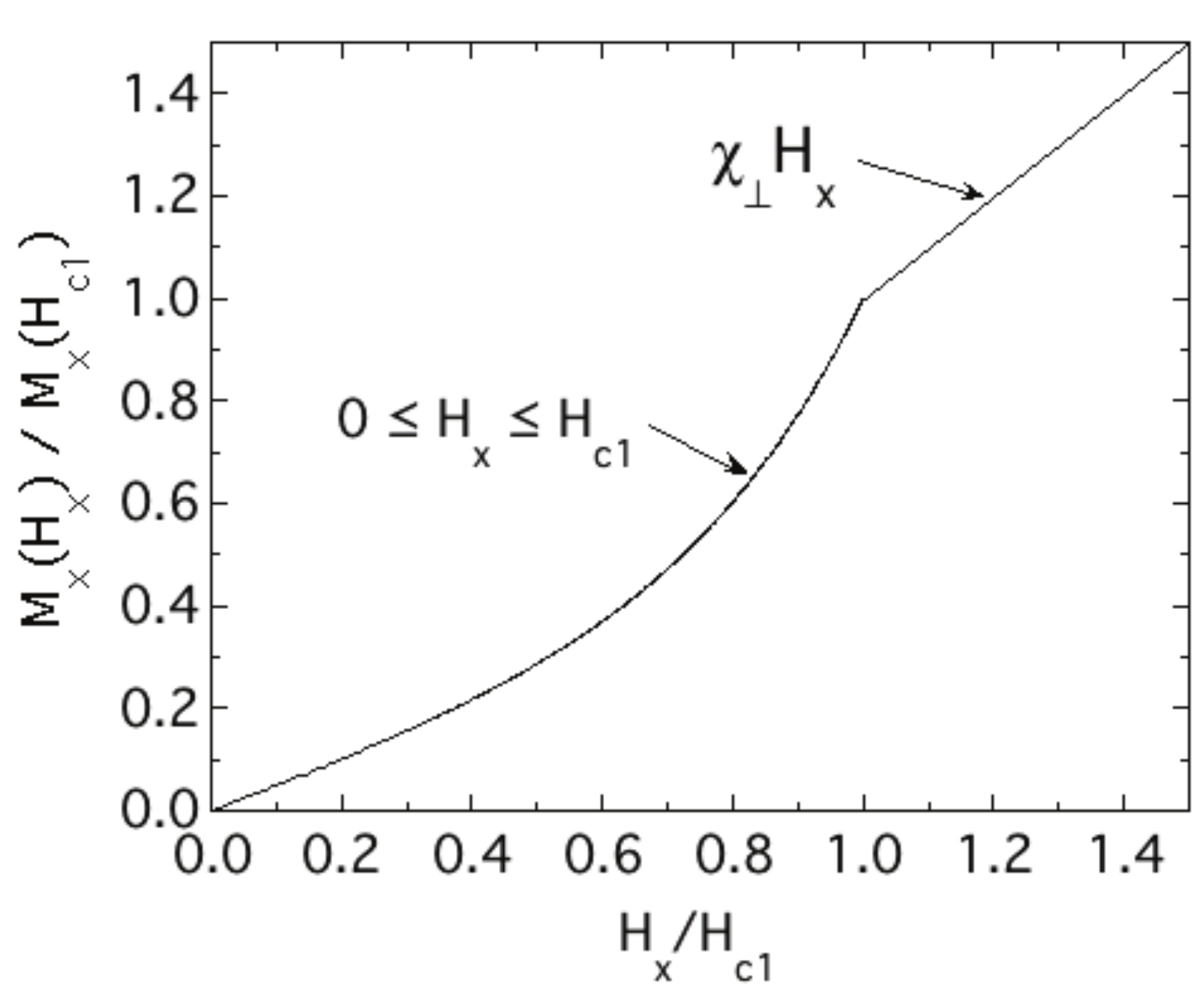}
\caption {Theoretical magnetization ratio $M_x(H_x)/M_x(H_{\rm c1})$ vs the field ratio $H_x/H_{\rm c1}$.  Above $H_{\rm c1}$, this ratio vs  $H_x/H_{\rm c1}$ is linear, which continues up to the critical field ratio $H^{\rm c}_{ab}/H_{\rm c1}$, above which the ratio  remains constant.  Here $\chi_\perp$ is the magnetic susceptibility along the $x$~axis in the $ab$~plane that is perpendicular to the moments at $H_x=H_{\rm c1}$ in Fig.~\ref{Fig:TrigDomains}(b).}
\label{Fig:Mx_Mx(Hc1)}
\end{figure}

When $H_{\rm c1}\leq H_x\leq H^{\rm c}_{ab}$, where $H^{\rm c}_{ab}$ is the critical field at which $M_x\equiv M_{ab}$ reaches saturation at $7\,\mu_{\rm B}$/Eu, the magnetization is due to canting of the moments towards $H_x$ in each domain as illustrated in Fig.~\ref{Fig:TrigDomains}(c).  In this regime, $M_x\propto H_x$ as seen at the higher fields in Fig.~\ref{Fig:EuMg2(Bi,Sb)2_MH}.  Combining the previous results, a plot of the calculated $M_x(H_x)/M_x(H_{\rm c1})$ versus $H_x/H_{\rm c1}$ is shown in Fig.~\ref{Fig:Mx_Mx(Hc1)}.  This plot looks similar to the experimental data for \emb\ and \ems\ in Fig.~\ref{Fig:EuMg2(Bi,Sb)2_MH}.  In the following section we obtain an expression for $K_3$ and its respective values for \emb\ and \ems.

%\clearpage

\section{\label{Fits} Fits to the low-$T$ magnetization data for E\lowercase{u}M\lowercase{g}$_2$B\lowercase{i}$_2$ and E\lowercase{u}M\lowercase{g}$_2$S\lowercase{b}$_2$}

\subsection{\label{Preface} Determining and interpreting the experimental value of $H_{\rm c1}$ from  $M(H_x)$ data at $T\ll T_{\rm N}$}

According to Fig.~\ref{Fig:Mx_Mx(Hc1)},  $H_{\rm c1}$ can be found from the experimental $M(H_x)$ data by calculating $dM(H_x)/dH_x$ versus $H_x$ and identifying the magnetic field at which the  peak occurs as $H_{\rm c1}$.  Then using the measured value of $H_{\rm c1}$   and the molar $\chi_\perp$ which is the molar magnetic susceptibility in the $ab$~plane at $T_{\rm N}$ within MFT, the value of the anisotropy constant $K_3$ can be calculated.  The value of $K_3$ per formula unit~(f.u.)~(per Eu atom in our case) is obtained from Eq.~(\ref{Eq:ChiPerpHc12}) as
\bea
K_3 = \frac{\chi_\perp H_{\rm c1}^2}{(9/2)N_{\rm A}} ,
\label{Eq:K3}
\eea
where $\chi_\perp=\chi_{ab}$, $N_{\rm A}$ is Avogadro's number, $\chi(T_{\rm N})$ is in cgs units of cm$^3$/mol, and $H_{\rm c1}$ is in cgs units of \mbox{Oe = G (Gauss).}  Using the cgs units of $\chi_{ab}(T_{\rm N})$ given by ${\rm cm^3/mol\,f.u.}$, the conversion factor ${\rm 1\,G^2 = 1\,erg/cm^3}$, and the units 1/mol of $N_{\rm A}$, the cgs units of $K_3$ are ${\rm ergs/f.u. = 1.602\times 10^{-12}~eV/f.u.}$

\subsection{\label{CalcAnisPars} Modeling the experimental $M_{ab}(H)$ data at $T=1.8$~K for \emb\ and \ems\ }

\begin{figure}[h]
\includegraphics [width=3.3in]{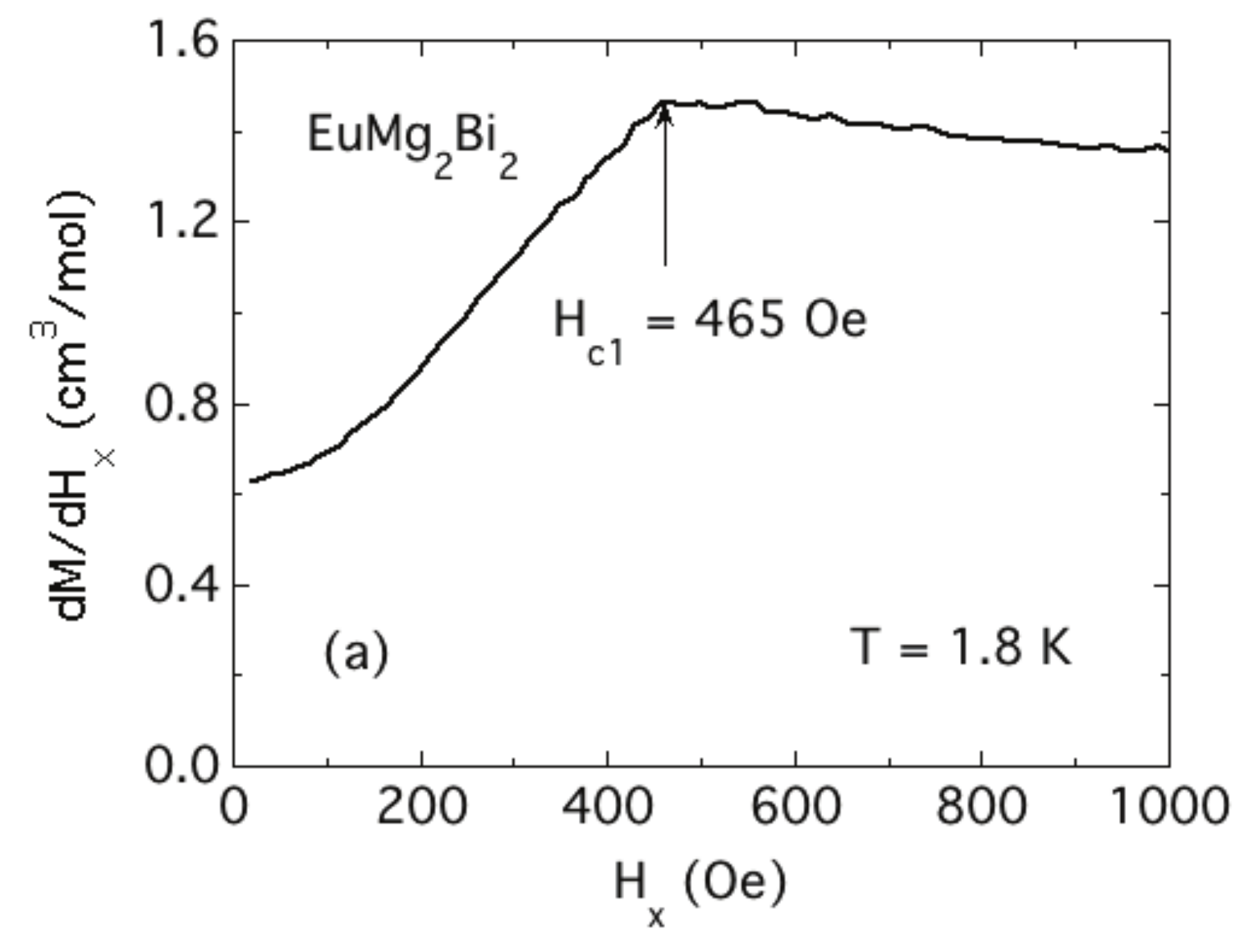}
\includegraphics [width=3.3in]{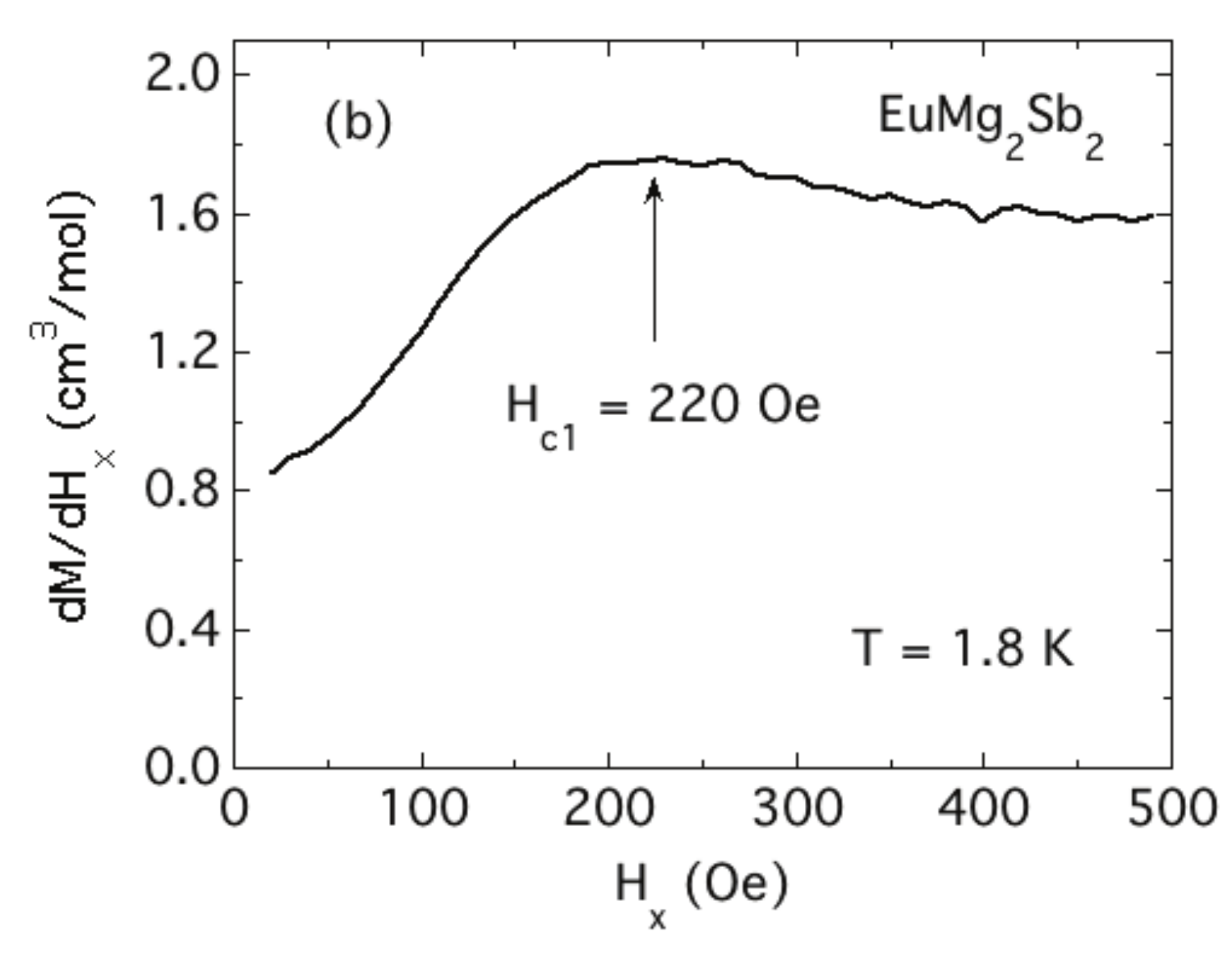}
\caption {Magnetic field derivative $dM/dH_x$ vs $H_x (=H_{ab})$ of the $M_x(H_x)$ data for \emb\ and \ems\ in Figs.~\ref{Fig:EuMg2(Bi,Sb)2_MH}(a) and \ref{Fig:EuMg2(Bi,Sb)2_MH}(b), respectively, in the low-field region from which the critical fields $H_{\rm c1} \approx 465$ and $\approx 220$~Oe are obtained, respectively.  The high-field behavior eventually asymptotes to the respective $ab$-plane magnetic susceptibilities \mbox{$\chi_{ab}(T_{\rm N}) = 1.31$} and  \mbox{$1.54~{\rm cm^3/mol}$}, respectively.}
\label{Fig:EuMg2Bi2dMdH}
\end{figure}

\begin{figure}[h]
\includegraphics [width=3.3in]{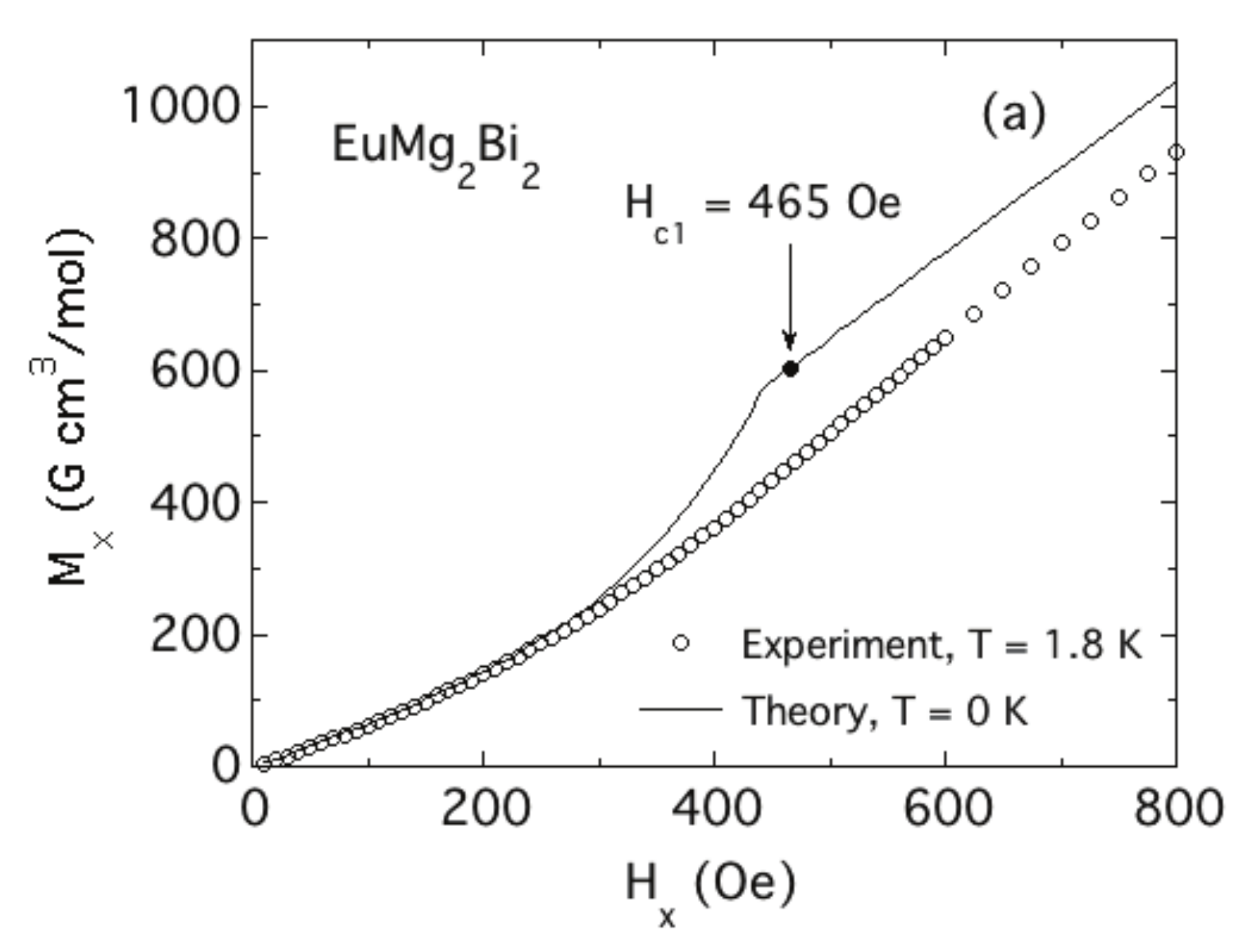}
\includegraphics [width=3.3in]{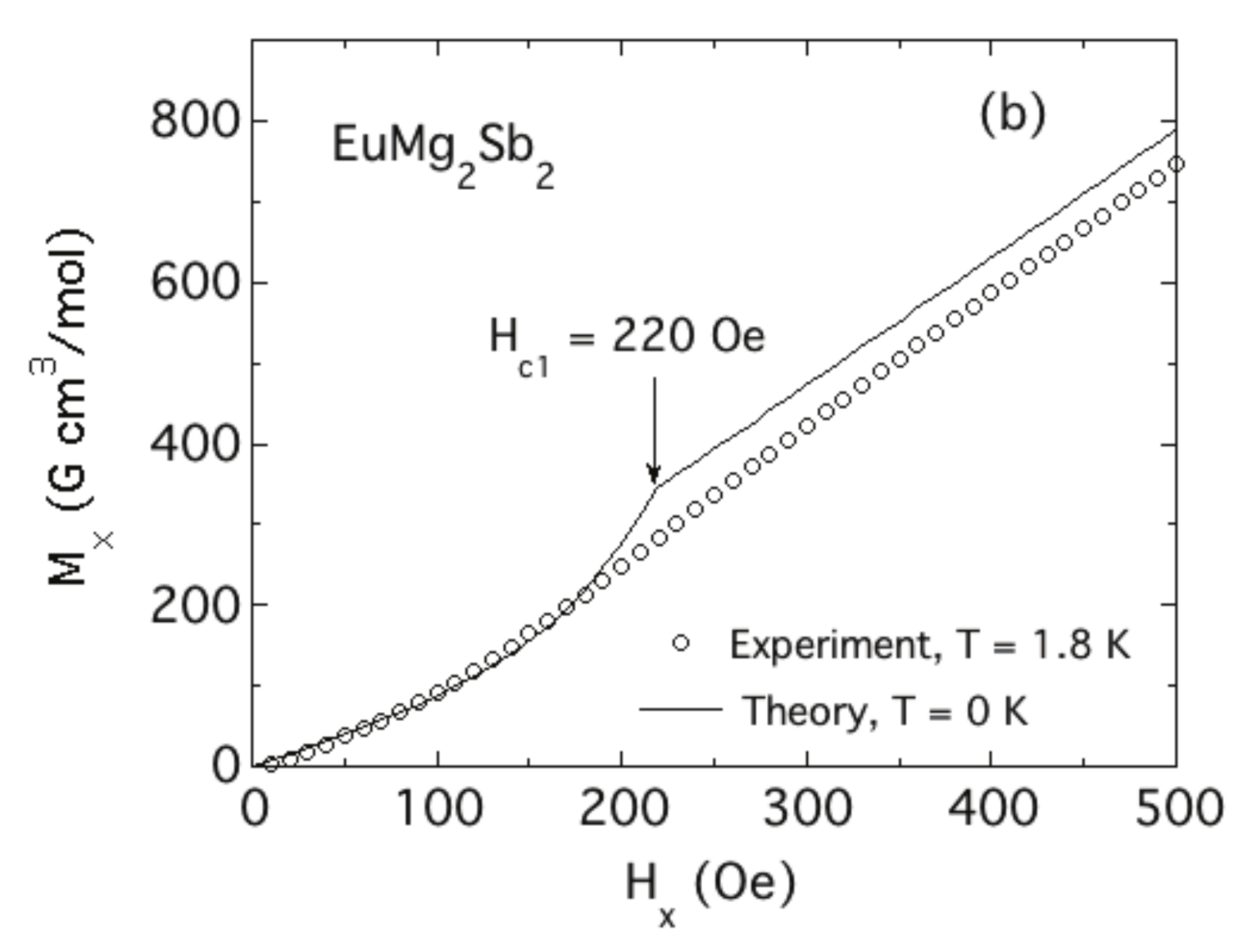}
\caption{Experimental magnetization $M_x=M_{ab}$ versus magnetic field~$H_x=H_{ab}$ at $T=1.8$~K (open circles) and the theoretical predictions for $T=0$~K (solid curve) for (a)~\emb\ with $H_{\rm c1} \approx 465$~Oe and (b)~\emb\ with $H_{\rm c1} \approx 220$~Oe. }
\label{Fig:MvsHx_data_fit_LowHx}
\end{figure}

Figures~\ref{Fig:EuMg2Bi2dMdH}(a) and \ref{Fig:EuMg2Bi2dMdH}(b) show the derivative $dM/dH_x$ vs $H_x$ in the $ab$~plane for \emb\ and \ems\ obtained from the  data in Figs.~\ref{Fig:EuMg2(Bi,Sb)2_MH}(a) and~\ref{Fig:EuMg2(Bi,Sb)2_MH}(b), respectively.  The data exhibit  maxima in $dM/dH_x$ which yield $H_{\rm c1}=465$~Oe for \emb\ and 220~Oe for \ems, respectively.  Then using the value $\chi_{ab}(T_{\rm N}) = 1.31$~cm$^{3}$/mol from Fig.~5(a) of Ref.~\cite{Pakhira2020} which is the same as the value of the high-field derivative at $H_{\rm c1} \approx  465$~Oe in Fig.~\ref{Fig:EuMg2Bi2dMdH}(a), the value of $K_3$ is obtained from Eq.~(\ref{Eq:K3}) for \emb.  Similarly, for \ems\  $\chi_{ab}(T_{\rm N}) = 1.58$~cm$^{3}$/mol and  $H_{\rm c1} \approx  220$~Oe. The results for the respective trigonal anisotropy constants are
\bse
\label{Eqs:K3Values}
\bea
K_3 &=& 1.0\times10^{-19}~{\rm erg/Eu} \quad ({\rm EuMg_2Bi_2}) \label{Eq:K3EMB}\\
&=& 6.5\times10^{-8}~{\rm eV/Eu},\nonumber\\
K_3 &=& 2.8\times10^{-20}~{\rm erg/Eu} \quad ({\rm EuMg_2Sb_2})\label{Eq:K3EMS}\\
&=& 1.8\times10^{-8}~{\rm eV/Eu}.\nonumber
\eea
\ese
Thus the trigonal anisotropy constant $K_3$ in \ems\ is significantly smaller than in \emb.

Comparisons of the experimental $M_{ab}(H)$ data at $T=1.8$~K for \emb\ in Fig.~\ref{Fig:EuMg2(Bi,Sb)2_MH}(a) and~\ems\ in Fig.~\ref{Fig:EuMg2(Bi,Sb)2_MH}(b) with the theoretical predictions at $T=0$~K using the experimental values of $H_{\rm c1}$ and $M_x(H_{\rm c1})$ are shown in Figs.~\ref{Fig:MvsHx_data_fit_LowHx}(a) and~\ref{Fig:MvsHx_data_fit_LowHx}(b), respectively. The theory accurately describes the experimental data below $H_{\rm c1}$ and the slope of the data above $H_{\rm c1}$, but the latter prediction is displaced upwards from the experimental data for both compounds.  We speculate that this difference arises from the finite temperature 1.8~K of the measurements, which is a significant fraction of $T_{\rm N} = 6.7$~K for \emb\ and $T_{\rm N} =8.0$~K for \ems, compared with $T=0$~K for the theory.

\subsection{\label{TheoryAnisValues} Theoretical anisotropy values}

\subsubsection{Magnetic Dipole and Critical-Field Anisotropies}

The energy of interaction $E_i$ of a magnetic moment $\mu_i$ due to the magnetic dipole interaction (MDI) with the other identical moments in a magnetically-ordered crystal is given by~\cite{Johnston2016}
\bse
\label{Eqs:MagDip}
\bea
E_{i\alpha} = -\epsilon \lambda_{k\alpha},
\label{Eq:Eialpha}
\eea
with
\bea
\epsilon = \frac{\mu^2}{2a^3}.\label{Eq:epsilon}
\eea
\ese
The $\lambda_{k\alpha}$ values are the eigenvalues of the dimensionless symmetric MDI tensor, where $k$ is the magnetic propagation vector and $\alpha$ is the ordered-moment axis~$\hat{\mu}$. The ordered magnetic moments are described by $\vec{\mu}_i = \mu\hat{\mu}_i$, and $a$ is the basal-plane lattice parameter of the respective crystal structure.

\paragraph{\emb} 

For $c/a = 1.644$~\cite{Pakhira2020} and the A-type AFM propagation vector (0,0,$\frac{1}{2}$)~r.l.u.~\cite{Pakhira2021}, the MDI tensor eigenvalues for the three Cartesian ordered-moment axes are \mbox{$\lambda_{[100]} = 5.515\,488\,367\,755\,699$,} \mbox{$\lambda_{[010]} = 5.515\,488\,367\,755\,807$} using high-precision calculations, and $\lambda_{[001]} = -11.0386$. The in-plane anisotropy energy $\Delta E$ per formula unit of \emb\ is given by
\bea
\Delta E = -\epsilon (\lambda_{[100]}-\lambda_{[010]}),
\eea
where the moment directions are in Cartesian coordinates with the $a$ axis designated by [100] and a direction perpendicular to it in the $ab$~plane by [010].  This would correspond to the amplitude of oscillation of the anisotropy energy in Fig.~\ref{Fig:AnisEnergyVsPhi}.   Using the values $a = 4.7724\times 10^{-8}$~cm and $\mu = 7~\mu_{\rm B}$~\cite{Pakhira2021, Pakhira2022}, we obtain $\Delta E = 3.00956\times10^{-18}$~{\rm eV}/Eu.  This value is a factor of $10^{10}$ smaller than estimated for $K_3$ in Eq.~(\ref{Eq:K3EMB}).  As a reference point, taking $\mu=7\,\mu_{\rm B}$ and $H_{\rm c1}=465$~Oe gives $\mu\,H_{\rm c1}=1.88\times 10^{-5}$~eV\@. Thus the in-plane anisotropy energy associated with in-plane locked-moment reorientation in the AFM domains in Fig.~\ref{Fig:TrigDomains}(a) must arise from a source other than the magnetic-dipole interaction.

The magnetic-dipole anisotropy energy difference between the observed $ab$~plane and $c$~axis ordering is
\bse
\label{EqsDeltaE}
\bea
\Delta E &=& -\epsilon (\lambda_{[100]}-\lambda_{[001]}) \\
&=&-0.2004~{\rm meV}.
\eea
The equivalent magnetic anisotropy field $\Delta H$ is
\bea
\Delta H = \frac{\Delta E}{\mu} = -4945~{\rm Oe}.
\label{Eq:DeltaH}
\eea
\ese
Thus, $ab$-plane ordering in the A-type AFM state of \emb\ is favored over $c$-axis ordering associated with magnetic-dipole interactions, as observed.  This is consistent with the critical fields $H^{\rm c}$ at $T=1.8$~K obtained from $M(H)$ isotherms $H_c^{\rm c}= 40(3)$~kOe and $H_{ab}^{\rm c}= 27.5(2)$~kOe~\cite{Pakhira2020}, yielding  $H_c^{\rm c}- H_{ab}^{\rm c}\approx12,000$~Oe.  This is because it is easier to cant the moments in each domain towards an $ab$-plane field than it is to cant them along the $c$~axis according to Eqs.~(\ref{EqsDeltaE}).  On the other hand, this difference in critical fields is significantly larger than the magnitude of $\Delta H$ in Eq.~(\ref{Eq:DeltaH}), suggesting the presence of a source of $\Delta H$ in addition to the magnetic-dipole interaction.

\paragraph{\ems} 

The corresponding magnetic-dipole calculations for \ems\ are given in the Appendix of Ref.~\cite{Pakhira2022}.  At 6.6~K, the lattice parameters are \mbox{$a=4.5431$~\AA,} \mbox{$c=7.6668$~\AA,} $c/a=1.6477$.   The results for A-type AFM ordering are 
\bse
\bea
\lambda_{[100]} &=& 5.519\quad {\rm (a~axis~ordering)}\\
\lambda_{[001]} &=& -11.038~~ {\rm (c~axis~ordering)}\\
\lambda_{[100]}-\lambda_{[001]} &=& 16.557\\
\Delta E &=& -0.2162~{\rm meV}\\
\Delta H &=& -5334~{\rm Oe}.
\eea
\ese
At $T=1.8$~K, the measured critical fields are \mbox{$H_c^{\rm c}= 34(1)$~kOe} and \mbox{$H_{ab}^{\rm c}= 26(1)$~kOe,} yielding \mbox{$H_c^{\rm c}- H_{ab}^{\rm c}\approx8.0$~kOe}, which indicates that the magnetic-dipole interaction is an important source of magnetic anisotropy in \ems, as in \emb.

The experimental results are summarized in Table~\ref{Tab.ExpData}.

%% Please fill in the text.  You may want to add theoretical details to the Experimental and Theoretical Details Section \ref{Details}.

\begin{table*}
\caption{\label{Tab.ExpData} Summary of experimental data for \emb\ and \ems.  Listed for each compound are the hexagonal lattice parameters $a$, $c$, and $c/a$, the value $\epsilon=\mu^2/(2a^3)$ in Eq.~(\ref{Eq:epsilon}) where the Eu moment is $\mu=7\,\mu_{\rm B}$, the N\'eel temperatures $T_{\rm N}$, and the critical fields $H_{\rm c1}$, $H^{\rm c}_{ab}$, and $H^{\rm c}_c$.  Except for $H_{\rm c1}$ obtained in the text, the values cited are from Refs.~\cite{Pakhira2020, Pakhira2022}.  The abbreviation RT stands for room temperature.}
\begin{ruledtabular}
\begin{tabular}{lcccccccc}	
Compound  			& $a$ 			& $c$ 			& $c/a$			&	$\epsilon$		& $T_{\rm N}$		&	$H_{\rm c1}$		& $H^{\rm c}_{ab}$	& $H^{\rm c}_c$	\\
 					&  (\AA)			& (\AA)			&				&	(meV)		& (K)				&  	(Oe)				& (kOe)	       		&  (kOe)	    		\\
\hline
\emb\ 				& 4.7724(3) (RT)	& 7.8483(5) (RT)	& 1.6445(2) (RT)	&	0.01210		& 	6.7(1)		&  	465(2)			& 27.5(2)			& 40(3)			 \\ 
\ems\ 				& 4.6861(3) (RT)	& 7.7231(5) (RT)	& 1.6481(2) (RT)	&				&	8.0(2)		&	220(10)			& 26(1)			& 34(1)			\\
				& 4.6531(5) (6.6 K)	& 7.6668(5) (6.6 K) 	& 1.6477(3) (6.6 K)	&	0.01306		&				&					&				&				\\
\end{tabular}
\end{ruledtabular}
\end{table*}

\begin{table*}
  \caption{\label{Tab.ThyData}
  Theoretical data for \emb\ and \ems.
  Here, {\bf k} is the propagation vector of the ordering in reciprocal-lattice units (r.l.u.)
  and $\vec{\alpha}$ is the collinear polarization of the ordered moments in real space.
For each magnetic configuration, the two arrows in the second column represent the spin orientations of two neighboring Eu layers.  
The magnetic-dipole energy (MDE) is calculated using the results in Ref.~\cite{Johnston2016}. 
The DFT energy calculated with spin-orbit coupling is also listed for both ferromagnetic (FM) and A-type antiferromagnetic (AFM) ordering, each for both $ab$-plane and $c$-axis moment alignment.
The total energy is calculated as the sum of MDE and DFT energy. 
The MDE per moment is calculated from the $\epsilon$ values in Table~\ref{Tab.ExpData} and the $\lambda_{k,\alpha}$ values using Eqs.~(\ref{Eqs:MagDip}) in the text.
All MDE calculations assume the Eu moment of $m_\text{Eu}=\SI{7}{\mub}$ and neglect the small induced moments on Bi/Sb and Mg sites.
For DFT energy and total energy, the ground-state values are chosen as reference zero.
Due to the negative sign in Eq.~(\ref{Eq:Eialpha}), the most probable magnetic structure in the second column for each compound due to MDE alone is the one with the largest positive value of $\lambda_{{\bf k},\alpha}$ in the sixth column and hence the most negative value in the eighth column.
The magnetic anisotropy energy (MAE) contains two contributions: magnetocrystalline anisotropy energy (MCAE) and magnetic-dipole anisotropy energy (MDAE).
They are listed for both A-type AFM and FM ordering as the corresponding energies of the $c$-axis moment alignment above those of the $ab$-plane moment alignment.
All energies listed here are in units of meV/Eu.
} \bgroup \def\arraystretch{1.2}
\begin{tabular*}{\linewidth}{c@{\extracolsep{\fill}}ccccccccccc}
  \hline \hline  
  Compound  & \,\,Configuration & Ordering & $\mathbf{k}$ (r.l.u.) & $\vec{\alpha}$ & $\lambda_{{\bf k},\alpha}$  & DFT Energy & MDE & Total Energy & MCAE & MDAE  & MAE \\ \hline %% \cline{6-6} \cline{7-7} \cline{8-8}
%%                                    &               &             &         &          & (meV/Eu)   & (meV/Eu)      & (meV/Eu)  & (meV/Eu) & (meV/Eu)  & (meV/Eu) \\ \hline      
  \multirow{4}{*}{EuMg$_2$Bi$_2$} &$\rightarrow\leftarrow$  & A-type AFM    & (0,0,1/2)   & [1 0 0] &   5.519  & 0          &    -0.0668    &       0   & 0.1255 &  0.2004  & 0.3259 \\ 
                                  &$\rightarrow\rightarrow$ & FM            & (0,0,0)     & [1 0 0] &   2.571  & 1.6195     &    -0.0311    &  1.6552   & 0.0205 &  0.0934  & 0.1139 \\ 
                                  &$\uparrow\downarrow$     & A-type AFM    & (0,0,1/2)   & [0 0 1] & -11.038  & 0.1255     &     0.1336    &  0.3259   &  &   &  \\ 
                                  &$\uparrow\uparrow$       & FM            & (0,0,0)     & [0 0 1] &  -5.142  & 1.6400     &     0.0623    &  1.7691   &  &   &  \\ \hline
  \multirow{4}{*}{EuMg$_2$Sb$_2$} &$\rightarrow\leftarrow$  & A-type AFM    & (0,0,1/2)   & [1 0 0] &   5.519  & 0          &    -0.0721    &       0   & 0.0380 &  0.2162  &  0.2542 \\ 
                                  &$\rightarrow\rightarrow$ & FM            & (0,0,0)     & [1 0 0] &   2.577  & 1.2775     &    -0.0337    &  1.3159   & 0.0265 &  0.1011  &  0.1276 \\ 
                                  &$\uparrow\downarrow$     & A-type AFM    & (0,0,1/2)   & [0 0 1] & -11.038  & 0.0380     &     0.1441    &  0.2542   &  &   &  \\ 
                                  &$\uparrow\uparrow$       & FM            & (0,0,0)     & [0 0 1] &  -5.155  & 1.3040     &     0.0674    &  1.4435   &  &   &  \\  \hline \hline
\end{tabular*}        
\egroup
\end{table*}

\subsubsection{Magnetocrystalline Anisotropy Energy (MCAE) from Density-Functional Theory}

\rtbl{Tab.ThyData} summarizes the calculated DFT energy, magnetic-dipole energy (MDE), and total energy for four different magnetic configurations in EuMg$_2$Bi$_2$ and EuMg$_2$Sb$_2$.
Two observations are consistent with experiments: (1) The A-type AFM with an in-plane spin orientation has the lowest total energies calculated in both compounds;
(2) \emb\  has a larger easy-plane anisotropy than \ems.
The magnetic anisotropy (MA, $K_\text{Total}$) contains two contributions: MDA and the SOC-originated MCA\@.
The Eu-$4f$ shell is half-filled in these compounds, resulting in a large spin moment of $\sim\SI{7}{\mub/Eu}$ and a negligible orbital moment.
Rotating the roughly spherical Eu-$4f$ charge cloud in a crystal field costs little energy, resulting in a relatively small MCA compared to typical rare-earth-based magnets with open $4f$-shell orbitals.
On the other hand, the large Eu spin moment gives considerable MDE and corresponding MDA. 
For example, for the ground-state A-type ordering, $K$ and $K_\text{D}$ are 0.1255 and 0.2004 meV/Eu, respectively, in \emb\  while the values are 0.038 and 0.2162~meV/Eu, respectively, in \emb.
The MDE and $\kd$ are similar in two compounds as they share similar lattice parameters and Eu magnetic on-site moments.
Thus, $\kd$ dominates the easy-plane MA for the A-type AFM ground state, especially in EuMg$_2$Sb$_2$, where the MCA is weaker.
On the other hand, the larger easy-plane MA in \emb\  is due to its larger MCA.

Besides the easy-plane anisotropy, we also calculated the in-plane MCAE by varying the spin orientation within the $ab$ plane; the obtained energy difference is smaller than \SI{2E-7}{eV/Eu}.  This result is consistent with Eqs.~(\ref{Eqs:K3Values}) which indicate that the in-plane anisotropy energies are $K_3 = 1.8$ and $6.5\times10^{-8}~{\rm eV/Eu}$ for  \ems\ and \emb, respectively.  
Thus an accurate estimation of the small in-plane MCAE is beyond the resolution of the present DFT approach.
As noted previously, the in-plane MDAE is negligible.

\begin{figure}[htb]
  \centering
  \begin{tabular}{c}
    \includegraphics[width=1.0\linewidth,clip]{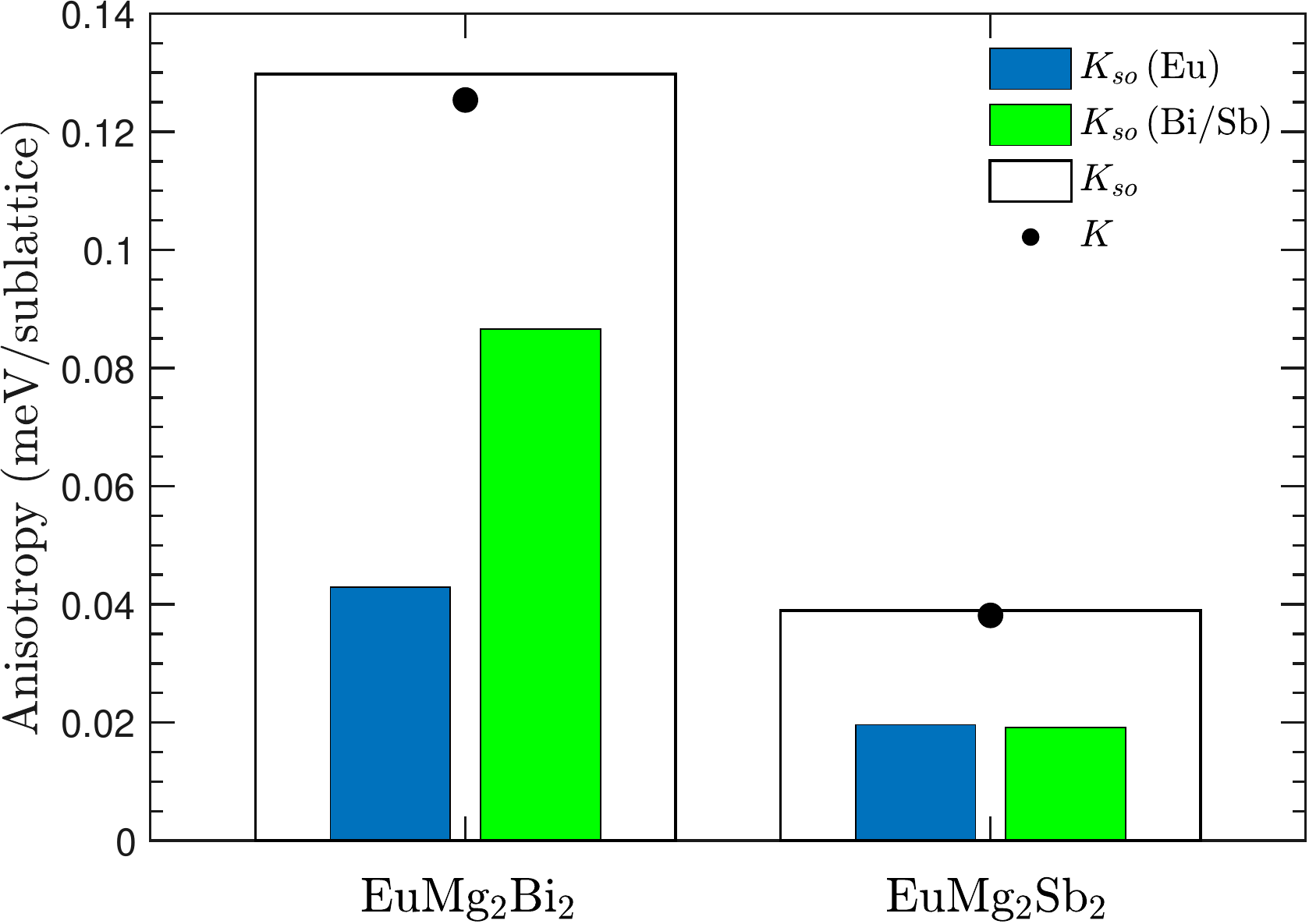} 
  \end{tabular}%   
  \caption{Total and sublattice-resolved anisotropy of spin-orbit-coupling energy $\kso$, defined as in \req{eq:kso}, in $A$-type AFM EuMg$_2$Bi$_2$ and EuMg$_2$Sb$_2$.
    Magnetocrystalline anisotropy energy (MCAE) $K$ is also shown (black dots) and agrees well with total $\kso$ (black box).
    The MCAE is calculated according to \req{eq:kc} using the total energies.
    The $\kso$ contribution from Mg sublattice $\kso$(Mg) is negligible and not shown. 
}
  \label{fig:mae}
\end{figure}

\paragraph{Origin of easy-plane MCA}
The larger easy-plane MCA in \emb\  than in \ems\  can be understood by resolving the anisotropy of SOC energy $\kso$ into sublattices~\cite{antropov2014ssc,ke2016prbA}.
Although the MCA is generally associated with the Eu spin's alignment, it also depends on the nature of the other constituent atoms in the compound. 
\rFig{fig:mae} shows the sublattice-resolved and total $\kso$, compared with MCAE~$K$\@.
The difference between the total $\kso$ and $K$ is within 4\% in both compounds.
Within perturbation theory, $\kso$ is proportional to the difference between spin-parallel and spin-flip
components of the orbital susceptibilities, vanishing in the nonmagnetic limit~\cite{ke2015prb}.
The strongly-magnetic Eu atoms spin-polarize the otherwise nonmagnetic Bi/Sb atoms, inducing sizable $\kso$ on the Bi and Sb sublattices.
The $\kso$(Bi) in \emb\ is larger than the $\kso$(Sb) in \ems, which is likely due to the larger Bi-$p$ SOC constant ($\sim\SI{2}{\eV}$) than that of Sb-$p$ ($\sim\SI{0.74}{\eV}$).
Furthermore, the $\kso$(Eu) is larger in \emb\ than in \ems.
On the other hand, the light Mg atom has a very small SOC constant and negligible $\kso$.
We note that such an MA mechanism that combines the strongly-magnetic $3d$ (or here $4f$) elements with the large-SOC heavy-$p$ elements is also responsible for the MA in many other systems, such as the two-dimensional van der Waals materials CrI$_3$, MnBi$_2$Te$_4$, and MnSb$_2$Te$_4$~\cite{gordon2021jpd,li2020prl,riberolles2021prb}.
Similarly, MnBi$_2$Te$_4$ has been found to have a much larger MCA than MnSb$_2$Te$_4$~\cite{li2020prl}.

\begin{table}[htbp]
  \caption{Coefficients ($C$) of the isotropic interlayer exchange $J_c$, anisotropic interlayer exchange $\gamma_c$, and single-ion anisotropy $A$ for the four different spin configurations when mapped onto the spin Hamiltonian defined in \req{eq:ham}.
  The four configurations and their energies are detailed in \rtbl{Tab.ThyData}.
  The three columns on the right in the present table represent the differences of the coefficients with respect to the ground state ($\rightarrow\leftarrow$), where $\Delta C =C-C^{\rightarrow\leftarrow}$.
}
\label{tbl:spin_coefficients}
\bgroup
\def\arraystretch{1.2}%  1 is the default, change whatever you need
\begin{tabular*}{\linewidth}{c @{\extracolsep{\fill}} rrrrcrrr}
  \hline \hline
\multirow{2}{*}{Configuration}  & \multicolumn{3}{c}{$C$}       &  & \multicolumn{3}{c}{$\Delta C$}   \\ \cline{2-4}  \cline{6-8}
                      & $J_c$ & $\gamma_c$ & $A$ &  & $J_c$ & $\gamma_c$ & $A$  \\ \hline
$\rightarrow\leftarrow$  &    -2 &  0  &   0 & &   0  &   0  &   0  \\
$\rightarrow\rightarrow$ &     2 &  0  &   0 & &   4  &   0  &   0  \\
$\uparrow\downarrow$     &    -2 & -2  &   1 & &   0  &  -2  &   1  \\
$\uparrow\uparrow$       &     2 &  2  &   1 & &   4  &   2  &   1  \\ \hline
\end{tabular*}\\
\egroup
\end{table}

\paragraph{Anisotropic exchange}

Since the $\kso$s of Bi/Sb contribute significantly to MCA and exist only when they are spin-polarized by neighboring Eu atoms, one may expect that the $\kso$(Bi/Sb) and total $K$ depend on the magnetic structure of the Eu moments.
Indeed, we found a large difference in the MCA between A-type and FM ordering in \emb.
This difference suggests that the MCA contains terms beyond single-ion terms when mapping the total energy into the effective magnetic Hamiltonian that includes only the Eu sites.
We map the energies of the four configurations listed in \rtbl{Tab.ThyData} into the Hamiltonian
\begin{equation}
  H=\sum_{i\ne j} J_{c}\,\hat{\bf e}_i \cdot \hat{\bf e}_j + \sum_{i\ne j} \gamma_{c} (\hat{\bf e}_i^z \hat{\bf e}_j^z) + \sum_{i} A_i (\hat{\bf e}_i^z)^{2},
  \label{eq:ham}
\end{equation}
where $J_c$ and $\gamma_c$ are the effective isotropic and anisotropic interlayer exchanges, respectively, $A$ is the single-ion anisotropy, and $\hat{\bf e}_i$ is the unit vector of the Eu magnetic moment at site~$i$.
The coefficients $C$ of $J_c$, $\gamma_c$, and $A$ for the four magnetic structures in Table~\ref{Tab.ThyData} are listed the \rtbl{tbl:spin_coefficients}.
The three effective interactions can be written as
\begin{equation}
\left(
\begin{array}{c}
 J_c \\ \gamma_c \\ K \\
\end{array}
\right)
=
\left(
\begin{array}{crc}
 4  &   0  &   0 \\
 0  &  -2  &   1 \\
 4  &   2  &   1 \\  
\end{array}
\right)^{-1}
\left(
\begin{array}{l}
 E_{\rightarrow\rightarrow}   \\
 E_{\uparrow\downarrow} \\
 E_{\uparrow\uparrow} \\  
\end{array}
\right).
\label{eq:jga}
\end{equation}
Using \req{eq:jga} and the DFT energy values listed in \rtbl{Tab.ThyData}, we obtain $J_c=0.405 (0.32)$~meV/Eu, $\gamma_c=-0.026 (-0.003)$~meV/Eu, and \mbox{$A=0.073~(0.032)$~meV/Eu} in \emb\ (\ems), indicating a significant anisotropic exchange $\gamma_c$ in \emb.

\section{\label{Conclusion} Concluding Remarks}

Our previous measurements of the in-plane magnetization $M(H_x)$ at low~$T$ (1.8~K) of trigonal \emb\ and \ems, which is below $T_{\rm N}$ of each compound, exhibited positive curvature for $H_x \lesssim 200$--500~Oe followed by a proportional behavior up to the respective critical field $H_{ab}^{\rm c}$~\cite{Pakhira2020, Pakhira2021}.  Here we formulated a theory which quantitatively explains these results.  Due to the threefold trigonal magnetic anisotropy of each compound shown in Fig.~\ref{Fig:TrigDomains}(a), the A-type AFM ordering leads to three collinear AFM domains at 120$^\circ$ to each other.  In an applied field $H_x$ in the $ab$ plane, the ordered moments in each domain rotate with increasing field to become approximately perpendicular to ${\bf H}_x$ at a critical field $H_{{\rm c}1}\approx 220$~Oe for \ems\ and $\approx 465$~Oe for \emb\ as illustrated in Fig.~\ref{Fig:TrigDomains}(b).  Then on further increasing $H_x$, the moments all tilt towards ${\bf H}_x$ in such a way that $M(H_x)$ is linear in $H_x$ as shown in Fig.~\ref{Fig:TrigDomains}(c) until all moments are parallel to the field at the critical field $H_{ab}^{\rm c}$ which is 26~kOe for \ems\ and 27.5~kOe for \emb.

Magnetic-dipole calculations showed that $ab$-plane ordered-moment alignment is strongly favored over \mbox{$c$-axis} alignment.  From the values of $H_{{\rm c}1}$, our theory allowed us to determine the in-plane trigonal anisotropy constant $K_3 = 1.8\times 10^{-8}$~eV/Eu for \ems\ and $K_3 = 6.5\times 10^{-8}$~eV/Eu for \emb.   The in-plane magnetic anisotropy due to magnetic dipole interactions is found to be far too small to account for the observed values of $K_3$, and hence must arise from an alternate mechanism.  However, the $K_3$ values are below the resolution of our DFT calculations.  On the other hand, our DFT calculations of the magnetocrystalline anisotropy energy including spin-orbit coupling showed XY anisotropy resulting in $ab$-plane magnetic ordering with A-type antiferromagnetic ordering favored over ferromagnetic ordering.  Furthermore, anisotropic exchange between the Eu spins was found to be significant in \emb.

An interesting avenue for future work is to calculate the temperature dependence of the in-plane magnetization $M(H_x)$ up to the respective $T_{\rm N}$ of \ems\ and~\emb\ for comparison with the corresponding data in Refs.~\cite{Pakhira2020, Pakhira2021}.  Another avenue is to investigate the low $ab$-plane field behavior of A-type-antiferromagnets with a layered tetragonal Eu structure and with the ordered moments in the $ab$~plane which would have twofold in-plane anisotropy instead of the threefold trigonal in-plane anisotropy in \ems\ and \emb.

\acknowledgments

We thank Robert McQueeney, Benjamin Ueland, and David Vaknin for helpful discussions.  This research was supported by the U.S. Department of Energy, Office of Basic Energy Sciences, Division of Materials Sciences and Engineering.  Ames Laboratory is operated for the \mbox{U.S. Department} of Energy by Iowa State University under Contract No.~DE-AC02-07CH11358.

\end{document}